\documentclass[reprint,aps,prd,twocolumn,notitlepage,nofootinbib,showpacs,preprintnumbers,superscriptaddress]{revtex4-1}
\usepackage[T1]{fontenc}
\usepackage[utf8]{inputenc}
\usepackage{bm}
\usepackage{amsmath}
\usepackage{amssymb}
\usepackage{esint}
\usepackage{color}
\usepackage{graphicx}
\PassOptionsToPackage{normalem}{ulem}
\usepackage{ulem}

\usepackage{times}
\usepackage{hyperref}
\usepackage{amsthm,bm,mathrsfs,stmaryrd,amsmath}

\newcommand{\be}{\begin{equation}}
\newcommand{\ee}{\end{equation}}
\newcommand{\ba}{\begin{eqnarray}}
\newcommand{\ea}{\end{eqnarray}}

\newcommand{\nn}{\nonumber\\}

\def\pa{\partial}

\def\a{\alpha}
\def\b{\beta}

\def\d{\delta}

\def\l{\lambda}
\def\L{\Lambda}
\def\m{\mu}
\def\n{\nu}

\def\r{\rho}

\def\s{\sigma}

\def\o{\omega}

\begin{document}


\title{Novel nonlinear kinetic terms for gravitons}



\author{Wenliang LI}
\email{lii.wenliang@gmail.com}
\affiliation{APC, Universit\'e Paris 7, CNRS/IN2P3, CEA/IRFU, Obs. de Paris, Sorbonne Paris Cit\'e, B\^atiment Condorcet, 
10 rue Alice Domon et L\'eonie Duquet, F-75205 Paris Cedex 13, France 
({UMR 7164 du CNRS})}
\affiliation{Crete Center for Theoretical Physics (CCTP) and Crete Center for Quantum Complexity and
Nanotechnology (CCQCN), Department of Physics, University of Crete, P.O. Box 2208, 71003, Heraklion, Greece}


\date{\today}

\begin{abstract}
A set of novel derivative terms for spin-2 fields are proposed. 
They are the wedge products of curvature two-forms and vielbeins. 
In this work, we investigate the properties of novel two-derivative terms in the context of bi-gravity. 
Based on a minisuperspace analysis, 
we identify a large class of bi-gravity models 
where the Boulware-Deser ghost could be absent.
We give a new perspective that 
Weyl Gravity and New Massive Gravity belong to this class of bi-gravity models 
involving novel derivative terms. 
In addition, we discuss the UV cut-off scales, dynamical symmetric conditions and novel higher-derivative terms. 

\end{abstract}


\maketitle

\section{Introduction.}
In the pursuit of going beyond Einstein, 
extensive efforts have been devoted to constructing gravitational theories 
that are different from general relativity. 
It is believed that the Einstein-Hilbert action with a cosmological constant 
is the only consistent nonlinear action for single massless spin-two field, 
so additional ingredients are necessary: 
Fierz-Pauli theory gives the graviton a mass term \cite{Fierz:1939ix}; 
Brans-Dicke theory introduces a scalar degree of freedom \cite{Brans:1961sx} ; 
Lovelock theory involves higher spacetime dimensions and higher derivative terms \cite{Lovelock:1971yv}.

Motivated by the unexpected accelerating expansion of the present universe \cite{accele}, 
more models were constructed in recent attempts, 
such as high derivative scalar theories \cite{horndeski} and 
nonlinear massive gravity \cite{deRham:2010kj}. 
One of the guiding principles is that 
a consistent model should be free of Ostrogradsky's ghost 
arising from higher order equations of motion. 
\\

Antisymmetrization is a universal element in these new models. 
Based on this ingredient, a general framework was developed for ghost-free
\footnote{Let us clarify that "ghost-free" in this framework means 
the building blocks are potentially free of Ostrogradsky's scalar ghost. 
In other words, the corresponding equations of motion for the scalar modes 
could be at most of second order. 
The scalar modes may come from 
the Helmholtz-Hodge decomposition of tensor fields. 
},
Lorentz-invariant, Lagrangian field theories \cite{gh-fr, Li:2015fxa}.  
In this framework, a set of novel kinetic terms for spin-2 fields were proposed
\be
\mathcal L_\text{kin} = R\big(E^{(1)}\big)\wedge E^{(2)}\wedge \dots\wedge E^{(d-1)},
\label{multi-kin}
\ee
where $d$ is the number of spacetime dimensions. 
They are the wedge products of geometric differential forms: 
a curvature two-form and several vielbeins. 
The vielbeins $E^{(k)}$ can be the same or different
\footnote{If all the vielbeins coincide, we have the standard Einstein-Hilbert kinetic term.}. 
Geometric intuition was used to construct these nonlinear terms. 

Along the line of Lovelock theory, 
we can build novel higher-derivative terms for spin-2 fields 
by introducing more curvature two-forms into the wedge products 
\be
R\big(E^{(1)}\big)\wedge \dots \wedge R\big(E^{(1)}\big)\wedge E^{(2)}\wedge \dots\wedge E^{(n)},
\ee
which is possible when spacetime has more than four dimensions. 
Lovelock terms correspond to the cases where all the vielbeins are the same.

If the wedge products do not involve derivative terms, 
they are nonlinear potential terms for spin-2 fields
\be
\mathcal L_\text{pot} =E^{(1)}\wedge \dots\wedge E^{(d)}.
\label{multi-pot}
\ee 
which include the cosmological constant term and 
other interacting potentials for spin-2 fields 
\cite{deRham:2010kj, Hassan:2011zd, Hinterbichler:2012cn, Bergshoeff:2013xma}
in the vielbein formulation \cite{Hinterbichler:2012cn}
\footnote{In 1970, J. Wess and B. Zumino already proposed 
to use vielbeins, rather than metrics, 
as the building blocks of  
the low energy effective potentials  
for spin-2 fields \cite{Zumino:1970tu}. 
However, it is still not fully clear 
why the vielbein formulation is more efficient 
in eliminating the Boulware-Deser ghost. 
}. 
These potential terms are usually free of the Boulware-Deser (BD) ghost
\footnote{In a concrete model, 
the BD ghost may be present even if the building blocks themselves are free of it. 
For example, the BD ghost is excited if loop-type interactions are introduced in the metric formulation \cite{Hinterbichler:2012cn}. 
Analogously, we find constraints on novel derivative terms. }. 
For simplicity, they are denoted by de Rham-Gabadadze-Tolley (dRGT) terms
\footnote{We refer to \cite{dRGT-reviews} for reviews of this subject. }. 
\\

The search for new derivative interactions was initiated in \cite{Folkerts:2011ev} 
where a new BD-ghost-free term in 4d was discovered
\be
{h_\m}^{[\m} {h_\n}^\n \pa_\r\pa^\r {h_\s}^{\s]}.\label{cubic}
\ee 

This cubic term can be thought of as a generalization of the perturbative Lovelock and dRGT terms 
and there are more possible terms in higher dimensions
\cite{Hinterbichler:2013eza}. 
It was also conjectured that they should admit nonlinear completions 
in parallel to the dRGT terms 
in the context of massive gravity \cite{Hinterbichler:2013eza}. 

Some of the novel kinetic terms in \eqref{multi-kin} are nonlinear, 
multi-gravity completions of this cubic term 
in terms of differential forms
\footnote{When the first version of this work appeared in arxiv, 
we were informed that similar nonlinear completions were studied before 
by K. Hinterbichler and R. A. Rosen.}. 
If we consider two spin-2 fields, impose the symmetric condition \cite{Deser:1974cy} 
and fix the second metric to Minkowski, 
the novel kinetic terms reduce to 
the two-derivative terms proposed in \cite{Kimura:2013ika}, 
which are nonlinear derivative terms for a massive graviton 
around a Minkowski background. 
They can also be obtained by dimensionally deconstructing 
the Gauss-Bonnet term \cite{deRham:2013tfa}. 
\\

In the discussions of \cite {Kimura:2013ika} and \cite{deRham:2013tfa}, 
only one metric is dynamical and 
the Boulware-Deser ghost was shown to be present. 
{\bf In this work, we will not make 
the single dynamical metric assumption, 
and the conclusion is different concerning the fate of the Boulware-Deser ghost. }
 
In \cite{deRham:2013tfa}, new kinetic interactions in 4d 
with a dynamical metric and a fixed Minkowski metric were investigated in detail. 
In the minisuperspace approximation, 
problematic $N^{-2}$ terms were found in the Lagrangians and 
the corresponding Hamiltonians are nonlinear in the lapse function $N$. 
This indicates the secondary constraint from the time derivative of $\pi_N=0$ is an equation for $N$
\footnote{The absence of an additional constraint was already found in 
\cite {Kimura:2013ika}, 
which should be present after a change of variables.
}. 
Then the dangerous, sixth degree of freedom remains dynamical. 
In 4d massive gravity, the sixth degree of freedom is ghost-like. 
It plagues a generic nonlinear completion of Fierz-Pauli theory 
and is known as the Boulware-Deser (BD) ghost \cite{Boulware:1973my}
\footnote{By an abuse of terminology, 
the sixth degree of freedom of a massless spin-2 field in 4d is also denoted by the Boulware-Deser (BD) ghost. }.
This ghost-like degree of freedom is eliminated in nonlinear massive gravity \cite{deRham:2010kj} thanks to the existence of a secondary constraint and an associated tertiary constraint \cite{Hassan:2011hr}.
Besides the minisuperspace discussion, 
an impressive no-go theorem was established in \cite{deRham:2013tfa} 
on new kinetic interactions for single dynamical metric models that are 
Lorentz-invariant and free of the BD ghost. 
\\
 
Inspired by the successful extensions of massive gravity \cite{deRham:2010kj} 
to bi-gravity \cite{Hassan:2011zd} and multi-gravity \cite{Hinterbichler:2012cn}, 
we would like to examine the bi-gravity models involving the novel derivative terms \eqref{multi-kin}. 
Given that pathologies were found in single dynamical metric models, 
it is very likely that the bi-gravity theories are sick as well. 
In fact, more recently, bi-gravity models with new kinetic interactions were studied 
in the first order formulation \cite{deRham:2015rxa}, where negative results were presented again
\footnote{To avoid confusion, let us emphasize that we consider second order formulation in this work, 
so spin connections are not independent. 
In other words, we assume the torsion-free condition.
}. Other obstructions were discussed in \cite{Matas:2015qxa} as well. 

Contrary to the single dynamical metric models, 
our analysis in section \ref{sec-mini} show that 
the sixth degree of freedom could be absent in some bi-gravity models 
constructed from \eqref{multi-kin}. 
But the price to pay is that 
one of the linearized kinetic terms has a wrong sign 
or at least one of them vanishes. 
In the former case, we encounter spin-2 ghosts, 
which can lead to tree-level non-unitarity upon quantization. 
In the latter case, the bi-gravity theories are strongly coupled 
due to the absence of kinetic terms. 
\\

The presence of a spin-2 ghost is a well-known feature of 
a generic model of higher derivative gravity. 
In this work, we propose a new viewpoint that, 
when the couplings to matter are not introduced,
a large type of bi-gravity models are equivalent to higher derivative gravity 
without Ostrogradsky's scalar ghost.  
They include Weyl gravity, 3d New Massive Gravity and 
some of their generalizations.

At the classical level, 
a spin-0 ghost is more dangerous than a spin-2 ghost. 
Usually, the Hamiltonian of a scalar ghost is unbounded from below, 
while that of a massless spin-2 field may simply vanish. 
In this sense, 
it is more crucial to eliminate the Boulware-Deser ghost. 
Furthermore, the Boulware-Deser ghost should be removed 
if a massive spin-2 field contains 
a correct number of dynamical degrees of freedom, 
which is at most 5 in four dimensions. 

Upon quantization, 
a spin-2 ghost will lead to tree-level non-unitarity when coupled to matter. 
Let us remind that a notorious problem in quantizing gravity is that 
the Einstein-Hilbert action is non-renormalizable \cite{non-ren}, 
which is very different from the other fundamental forces. 
By introducing higher order curvature terms 
(thus unitarity is sacrificed), 
one can improve the short-distance behaviour of the propagators and 
obtain a perturbative renormalizable theory for gravity \cite{Stelle}. 
Roughly speaking, the improved high energy behaviour are due to  
the relatively negative contributions from the ghost modes, 
which is analogous to the role of 
heavy ghost-like modes in Pauli-Villars regularization and 
superpartners in supersymmetric theories. 
\\

From a different perspective, 
we can consider a metric as an effective description of some microscopic physics. 
In an effective field theory (EFT) of gravity, 
higher order curvature terms are expected in a low energy expansion, 
because they are compatible with the symmetries \cite{Donoghue:1994dn}. 
Even if gravity itself is not quantized, 
they can be generated by quantum corrections from the matter. 
Therefore, some bi-gravity models with novel kinetic terms 
belong to a special class of effective field theories of gravity 
where Ostrogradsky's scalar ghost is removed. 

The cut-off scale of an EFT of gravity is usually associated with the Planck mass. 
However, if some of the high order curvature terms have unnaturally large coefficients, 
then the cut-off scale will be lowered. 
By eliminating the BD ghost, we could increase the cut-off scale set 
by large high order curvature corrections. 
\\

This paper is organized as follows. 
In section \ref{sec-kin-bi}, we give the precise formulation of the novel kinetic terms for 4d bi-gravity models. 
In section \ref{sec-mini}, we carry out a minisuperspace analysis to identify the candidate theories 
that are free of the dangerous, sixth degrees of freedom. 
In section \ref{sec-redef}, we perform a field redefinition 
to obtain more explicit expressions of the novel kinetic terms. 
In section \ref{sec-L2}, we focus on the novel kinetic term $\mathcal L_2{}^\text{kin}$ 
and discuss its relation to higher curvature gravity. 
In section \ref{sec-lin}, we linearize the nonlinear models and 
diagonalize the quadratic actions.
In section \ref{sec-s2gh}, the issue of spin-2 ghost is discussed. 
In section \ref{sec-cutoff}, we examine the cut-off scales of the effective field theories of gravity 
involving novel kinetic terms.  
In section \ref{sec-sym}, we explain how to obtain the symmetric condition from the equations of motion. 
In section \ref{sec-highder}, higher derivative generalizations are discussed and 
a general argument for the absence of the BD ghost is presented. 
In section \ref{sec-sum}, we summarize our results and discuss their implications.

\section{Novel kinetic terms for bi-gravity}\label{sec-kin-bi}

To be more specific, we mainly consider four dimensional spacetime and models with two vielbeins/metrics. 
There are six possible nonlinear kinetic terms
\ba
\mathcal L_1{}^\text{kin} =&\, R(E)\wedge E\wedge E,\label{L1}\\ 
\mathcal L_2{}^\text{kin} =&\, R(E)\wedge E\wedge F,\label{L2}\\
\mathcal L_3{}^\text{kin} =&\, R(E)\wedge F\wedge F\label{L3},
\ea
\ba
\mathcal L_4{}^\text{kin} =&\, R(F)\wedge F\wedge F,\label{L4}\\ 
\mathcal L_5{}^\text{kin} =&\, R(F)\wedge F\wedge E,\label{L5}\\
\mathcal L_6{}^\text{kin} =&\, R(F)\wedge E\wedge E,\label{L6}
\ea
where $R(E),\, R(F)$ are curvature two-forms
\ba
R(E)=d\,\o^E+\o^E\wedge\o^E,\\
R(F)=d\,\o^F+\o^F\wedge\o^F.
\ea 

Both $E$ and $F$ are dynamical vielbeins.  
$\o^E$ and $\o^F$ are the spin connections compatible with $E$ and $F$ respectively
\ba
D^E E=d\,E+\o^E\wedge E=0,\\
D^F F=d\,F+\o^F\wedge F=0.
\ea

Notice that one of the vielbeins in 
$\mathcal L_2{}^\text{kin}$ and $\mathcal L_4{}^\text{kin}$ 
are Lagrange multipliers. 
Another interesting observation is that 
$\mathcal L_3{}^\text{kin}$ and $\mathcal L_6{}^\text{kin}$ 
can be thought of as the Palatini formulation of the Einstein-Hilbert term, 
where the spin-connections are expressed in terms of the associated vielbeins
\footnote{The difference is that varying the action with respect to the vielbein in the curvature two-form will 
give rise to second order equations. 
``Torsion-free condition" is not the only solution, 
so $E$ and $F$ are not necessarily proportional to each other. }.

We also have five potential terms
\ba
\mathcal L_1{}^{\text{pot}}&=&E\wedge E \wedge E\wedge E, 
\label{bi-gr-pot-1}
\\
\mathcal L_2{}^{\text{pot}}&=&E\wedge E \wedge E\wedge F, 
\label{bi-gr-pot-2}
\\
\mathcal L_3{}^{\text{pot}}&=&E\wedge E \wedge F\wedge F, 
\label{bi-gr-pot-3}
\\
\mathcal L_4{}^{\text{pot}}&=&E\wedge F \wedge F\wedge F, 
\label{bi-gr-pot-4}
\\
\mathcal L_5{}^{\text{pot}}&=&F\wedge F \wedge F\wedge F. 
\label{bi-gr-pot-5}
\ea

When we discuss other dimensions, 
the subscripts $n$ in $\mathcal L_n$ means that 
the number of $F$ vielbeins in a wedge product is $(n-1)$. 
The two vielbeins ${E_\mu}^A$ and ${F_\mu}^A$ are related to 
two metrics $g_{\m\n}$ and $f_{\m\n}$
\be
g_{\m\n}={E_\m}^A{E_\n}^B \eta_{AB},\quad
f_{\m\n}={F_\m}^A{F_\n}^B \eta_{AB}.\label{metric-def}
\ee
The nonlinear kinetic terms $\mathcal L_1$ and $\mathcal L_4$ are 
the standard Einstein-Hilbert kinetic terms, 
while $\mathcal L_2,\,\mathcal L_3,\,\mathcal L_5,\,\mathcal L_6$
 are novel kinetic terms for spin-2 fields. 

To simplify our notation, the Levi-Civita symbol $\epsilon_{^{ABCD}}$ is not written explicitly 
in a wedge product. 
In terms of the components, 
the bi-gravity kinetic terms (\ref{L1}-\ref{L6}) 
and the bi-gravity potential terms (\ref{bi-gr-pot-1}-\ref{bi-gr-pot-5}) are
\ba
\mathcal L_1{}^\text{kin} =&\,\d_{^{ABCD}}^{\m\n\r\s}\, {R(E)_{\m\n}}^{AB} {E_\r}^C {E_\s}^D d^4x,\label{exp-1}
\\
\mathcal L_2{}^\text{kin} =&\,\d_{^{ABCD}}^{\m\n\r\s}\, {R(E)_{\m\n}}^{AB} {E_\r}^C {F_\s}^D d^4x,\label{exp-2}
\\
\mathcal L_3{}^\text{kin} =&\,\d_{^{ABCD}}^{\m\n\r\s}\, {R(E)_{\m\n}}^{AB} {F_\r}^C {F_\s}^D d^4x,\label{exp-3}
\ea
\ba
\mathcal L_4{}^\text{kin} =&\,\d_{^{ABCD}}^{\m\n\r\s}\, {R(F)_{\m\n}}^{AB} {F_\r}^C {F_\s}^D d^4x,\label{exp-4}
\\
\mathcal L_5{}^\text{kin} =&\,\d_{^{ABCD}}^{\m\n\r\s}\, {R(F)_{\m\n}}^{AB} {F_\r}^C {E_\s}^D d^4x,\label{exp-5}
\\
\mathcal L_6{}^\text{kin} =&\,\d_{^{ABCD}}^{\m\n\r\s}\, {R(F)_{\m\n}}^{AB} {E_\r}^C {E_\s}^D d^4x,\label{exp-6}
\ea
and
\ba
\mathcal L_1{}^\text{pot} =&\,\d_{^{ABCD}}^{\m\n\r\s}\,  {E_\m}^A {E_\n}^B {E_\r}^C {E_\s}^D d^4x,\label{exp-pot-1}
\\
\mathcal L_2{}^\text{pot} =&\,\d_{^{ABCD}}^{\m\n\r\s}\,  {E_\m}^A {E_\n}^B {E_\r}^C {F_\s}^D d^4x,\label{exp-pot-2}
\\
\mathcal L_3{}^\text{pot} =&\,\d_{^{ABCD}}^{\m\n\r\s}\,  {E_\m}^A {E_\n}^B {F_\r}^C {F_\s}^D d^4x,\label{exp-pot-3}
\\
\mathcal L_4{}^\text{pot} =&\,\d_{^{ABCD}}^{\m\n\r\s}\,  {E_\m}^A {F_\n}^B {F_\r}^C {F_\s}^D d^4x,\label{exp-pot-4}
\\
\mathcal L_5{}^\text{pot} =&\,\d_{^{ABCD}}^{\m\n\r\s}\,  {F_\m}^A {F_\n}^B {F_\r}^C {F_\s}^D d^4x,\label{exp-pot-5}
\ea

where ${R_{\m\n}}^{AB}$ are the components of the curvature two-forms. 
The antisymmetric Kronecker delta or the generalized Kronecker delta 
is an antisymmetric product of the Kronecker deltas
\be
\delta_{^{ABCD}}^{\m\n\r\s}={\d_{_{A}}}^{[\m} {\d_{_{B}}}^\n {\d_{_{C}}}^\r {\d_{_{D}}}^{\s]},
\ee
where the antisymmetrization $[\dots]$ is not normalized.
The Planck mass is not written explicitly. 
Greek letters $\m,\,\n,\,\r,\,\s,\dots$ indicate external spacetime indices and 
capital Latin letters $A,\,B,\,C,\,D,\dots$ denote internal Lorentz indices. \\

To minimalize the numbers of dynamical degrees of freedom, 
we impose the symmetric condition or 
the Deser-van Nieuwenhuizen condition \cite{Deser:1974cy}. 
In section \ref{sec-sym}, we discuss how the symmetric condition 
originates in equations of motion. 

\section{Minisuperspace analysis}\label{sec-mini}
In this section, we study the minisuperspace approximation of 
the bi-gravity models constructed from (\ref{L1}-\ref{L6}). 
The minisuperspace analysis is 
a simple test of the ghost-free condition 
before investigating the fully nonlinear structure. 
Although it is not sufficient to prove healthiness, 
it is very efficient in detecting pathologies. 
For example, it was used in \cite{Nomura:2012xr}  
to show that loop-type interactions of multi-gravity 
in the metric formulation can excite the BD ghost.

In \cite{deRham:2013tfa}, 
new kinetic interactions with single dynamical metric were ruled out, 
because their Hamiltonians are not linear in the lapse function 
in the minisuperspace approximation, 
which indicates the presence of the BD-ghost. 
Despite the failure of single dynamical metric models, 
a large class of bi-gravity theories do satisfy the criterion that 
the minisuperspace Hamiltonians are linear in lapse functions, 
as we discuss below. 
 
Now we start the minisuperspace analysis.  
The two metrics in the minisuperspace approximation are diagonal
\ba
ds_1^2=g^E_{\m\n}\,dx^\m dx^\n=-(N_1)^2dt^2+e^{2A}dx^2,
\label{mss-1}
\\
ds_2^2=g^F_{\m\n}\,dx^\m dx^\n=-(N_2)^2dt^2+e^{2B}dx^2.
\ea
where the metric components are functions of time
\be
N_1=N_1(t),\quad
N_2=N_2(t),\quad
A=A(t),\quad
B=B(t).
\ee

The corresponding symmetric vielbeins are
\be
{E_{\m}}^A
=\begin{pmatrix}
    N_1      & 0   \\
    0       &\quad e^A\, {\d_i}^j 
\end{pmatrix},
\quad
{F_{\m}}^A
=\begin{pmatrix}
    N_2      & 0   \\
    0       &\quad e^B\, {\d_i}^j 
\end{pmatrix}.
\ee

Let us consider a linear combination of 
$\mathcal L_1{}^\text{kin},\dots,\, \mathcal L_6{}^\text{kin}$,
so the Lagrangian reads
\ba
\mathcal L&=&
a_1\,\mathcal L_1{}^\text{kin}+
a_2\,\mathcal L_2{}^\text{kin}+
a_3\,\mathcal L_3{}^\text{kin}
\nn&&
+
b_1\,\mathcal L_4{}^\text{kin}+
b_2\,\mathcal L_5{}^\text{kin}+
b_3\,\mathcal L_6{}^\text{kin}.
\ea

In the minisuperspace approximation, 
it becomes
\ba
\mathcal L_{\text{mini}}
&=&
a_1 \frac {1} {N_1}(\dot A)^2e^{3A}
+b_1\frac {1} {N_2}(\dot B)^2e^{3B}
\nn&&
+
a_2\frac {1}{N_1}\left(\dot A^2+2 \dot A\dot B
-\frac {N_2}{N_1}\dot A^2 e^{A-B}\right)e^{2A+B}
\nn&&
+a_3
\frac 1 {N_1}\left(2\dot A \dot B 
-\frac {N_2}{N_1}\dot A^2 e^{A-B}\right)e^{A+2B}
\nn&&
+b_2\frac {1}{N_2}\left(\dot B^2+2\dot A\dot B
-\frac {N_1}{N_2}\dot B^2 e^{B-A}\right)e^{A+2B}
\nn&&
+b_3\frac 1 {N_2}\left(2\dot A \dot B 
-\frac {N_1}{N_2}\dot B^2 e^{B-A}\right)e^{2A+B},\label{mini}
\ea
where some normalization factors are inserted 
to simplify the expression of $\mathcal L_\text{mini}$. 
Time derivatives are denoted by dots
\be
\frac {d} {dt}A=\dot A,
\quad
\frac {d}{dt}B=\dot B.
\ee 
Integration by parts is performed in order to eliminate
\be
\dot N_1, \quad \dot N_2,\quad\ddot A,\quad\ddot B.
\ee

The conjugate momenta can be derived from the minisuperspace Lagrangian
\be
P_A=\frac {\pa \mathcal L_{\text{mini}}}{\pa \dot A},\quad
P_B=\frac {\pa \mathcal L_{\text{mini}}}{\pa \dot B},
\ee
\be
P_{N_1}=P_{N_2}=0,
\ee
where the second line are primary constraints. 
Then we derive the minisuperspace Hamiltonian 
by the Legendre transform
\be
\mathcal H_{\text{mini}}=\dot A P_A+\dot B P_B-\mathcal L_{\text{mini}}.
\ee

Using the relations between momenta and velocities, 
one can express the Hamiltonian $\mathcal H_{\text{mini}}$ in terms of
\be
N_1,\, N_2,\, A,\, B,\,P_A,\, P_B.
\ee 
The explicit expression of $\mathcal H_{\text{mini}}$ is a fraction
\be
\mathcal H_\text{mini}=\frac{\mathcal H_\text{n}}{4\mathcal H_\text{d}},
\ee
where the numerator $H_\text{n}$ and the denominator $H_\text{d}$ are
\ba
\mathcal H_\text{n}
&=&(b_3e^A+b_2 e^B)\,e^{-2A}\,P_A^2\,(N_1)^3
\nn&&
+(a_2e^A+a_3 e^B)\,e^{-2B}\,P_B^2\,(N_2)^3
\nn&&
+2(b_3 e^{A}+b_2 e^{B})\,e^{-(A+B)}\,P_AP_B\,(N_1)^2N_2
\nn&&
+2(a_2 e^{A}+a_3 e^{B})\,e^{-(A+B)}\,P_AP_B\,N_1(N_2)^2
\nn&&
-(b_2e^A+b_1 e^B)\,e^{-2A}\,P_A^2\, (N_1)^2 N_2
\nn&&
-(a_1e^A+a_2 e^B)\,e^{-2B}\,P_B^2 \,N_1 (N_2)^2,
\ea
\ba
\mathcal H_d&=&
(b_3 e^A + b_2 e^B) \big[(a_1 + b_3) e^A + (a_2 + b_2) e^B\big] (N_1)^2
\nn&&
+(a_2 e^A + a_3 e^B) \big[(a_2 + b_2) e^A + (a_3 + b_1) e^ B\big] (N_2)^2
\nn&&
-\big[(a_1 b_2 - a_2 b_3) e^{2 A} 
+ (a_2 b_1 - a_3 b_2) e^{2 B} 
\nn&&
\qquad+ (a_1 b_1 - a_3 b_3) e^{A + B}\big] N_1 N_2.
\ea

We require $\mathcal H_\text{mini}$ to be linear in $N_1$ and $N_2$, 
so $N_1$ and $N_2$ are Lagrange multipliers. 
Then the secondary constraints 
\be
\dot P_{N_1}=\{P_{N_1},\,\mathcal H_\text{mini}\}\approx 0,
\ee
\be
\dot P_{N_2}=\{P_{N_2},\,\mathcal H_\text{mini}\}\approx 0
\ee
are equations for the canonical variables 
and could remove the scalar modes 
related to the BD ghost
\footnote{Strictly speaking, 
we should use the total Hamiltonian which contains the primary constraints to compute the time derivative. }. 
The Poisson bracket is defined as
\be
\{\a,\, \b\}=\sum_{q=A,\,B,\,N_1,\,N_2}\left(\frac {\pa \a}{\pa q}\frac {\pa \b}{\pa P_q}-\frac {\pa \a}{\pa P_q}\frac {\pa \b}{\pa q}\right)
\ee

The numerator $H_\text{n}$ and the denominator $H_\text{d}$ 
are polynomials of degree three and two in $N_1$ and $N_2$. 
To satisfy the requirement that lapse functions are Lagrange multipliers, 
$H_\text{d}$ should be a factor of $H_\text{n}$. 
This is true when only one of the three monomials in $H_d$
has non-zero coefficient, 
which indicates two classes of bi-gravity models:
\begin{itemize}
\item
The first class is
\be
a_2=a_3=b_2=b_3=0,\label{1-class-bigra}
\ee
where the Lagrangian contains two Einstein-Hilbert terms 
and the minisuperspace Hamiltonian reads
\be
\quad
\qquad
\mathcal H^{\text{I}}_\text{mini}=N_1\left(\frac {e^{-3A}} {4a_1}   P_A^2\right)
+N_2\left(\frac {e^{-3B}} {4b_1}   P_B^2\right).
\ee
One can introduce the cosmological constant terms (\ref{bi-gr-pot-1},\,\ref{bi-gr-pot-5}).  
The interactions between the two metrics are 
through the nonlinear potential terms  
(\ref{bi-gr-pot-2},\,\ref{bi-gr-pot-3},\,\ref{bi-gr-pot-4}). 
The minisuperspace Hamiltonian is still linear in the lapse functions. 

Since we have two Planck masses in front of two Einstein-Hilbert terms, 
one can take the limit where one of them goes to infinity. 
In this decoupling limit, 
a bi-gravity model reduces to
that of single dynamical metric with a fixed metric.

\item
The second class is
\be
a_1=a_2=a_3=0
\ee
or 
\be
b_1=b_2=b_3=0\label{2cl-2}.
\ee
The bi-gravity models in the second class contain at most one Einstein-Hilbert kinetic term
(no EH term if $a_1=b_1=0$), 
so one can not take the decoupling limit that fixes one of the metrics
\footnote{There exists another single metric limit 
\be
E-F=H/ \l, \quad \l\rightarrow \infty,
\ee
where the novel kinetic terms reduce to the Einstein-Hilbert term.
} 
\footnote{The failure of obtaining nonlinear partially massless gravity 
from a consistent truncation of conformal gravity \cite{Deser:2012qg} 
is related to the absence of this decoupling limit.}
\footnote{The fact that only certain combinations of kinetic terms are allowed is analogous to 
the scalar-tensor theories discussed \cite{Langlois:2015cwa}, 
where the degeneracy conditions can break down for some combinations of degenerate Lagrangians. }.
Therefore, they are not ruled out 
by the no-go theorems in \cite{deRham:2013tfa}. 
We focus on the case of \eqref{2cl-2} in the discussions below.

The minisuperspace Hamiltonian with \eqref{2cl-2} reads
\ba
\mathcal H^{\text{II}}_\text{mini}&=&N_1\left[\frac {e^{-(A+B)}}{2(a_2\,e^A+a_3\, e^B)}P_A P_B\right.
\nn&&\qquad
\left.\quad-\frac{(a_1\, e^A+a_2 \,e^B)e^{-2B}}{4(a_2\,e^A+a_3\,e^B)^2} P_B^2\right]
\nn&&
+N_2\left[\frac {e^{-2B}}{4(a_2\,e^A+a_3\, e^B)} P_B^2\right],
\ea
where we assume $a_2$ and $a_3$ are not zero at the same time. 

We can introduce the potential terms $\mathcal L_1{}^\text{pot},\dots,\mathcal L_5{}^\text{pot}$. 
The lapse functions will still be Lagrange multipliers in the Hamiltonians. 

From the holographic point of view, 
the diagonal diffeomorphism invariance is fundamental 
in bi-gravity and multi-gravity theories \cite{KN}. 
In the context of the AdS/CFT correspondence, 
a conformal field theory with conserved stress tensor is dual to 
a diffeomorphism invariant theory. 
The massive gravitons 
(or more general the spin-2 fields without gauge invariance) 
correspond to spin-2 operators that are not conserved. 
From this perspective, 
a massive gravity theory should always admit enhancement 
to a diffeomorphism invariant theory 
by turning on the conserved stress tensor in the boundary field theory.  
However, the converse is less justified. 
A bi-gravity or a multi-gravity theory may not have a decoupling limit 
that breaks the diagonal diffeomorphism symmetry, 
which indicates the conserved stress tensor decouples. 
\end{itemize}

According to the minisuperspace Hamiltonians, 
there are two classes of bi-gravity models 
that are potentially free of the Boulware-Deser ghost. 
The first class of bi-gravity theories were proposed 
in \cite{Hassan:2011zd} 
by promoting the fixed metric 
in consistent non-linear massive gravity \cite{deRham:2010kj} . 
Only the standard Einstein-Hilbert kinetic terms are used. 
The absence of the BD ghost was proved in \cite{Hassan:2011hr}. 

The second class of bi-gravity models \eqref{2cl-2} are in 
a different region of the parameter space where novel kinetic terms are used. 
Let us investigate them in more detail. 
We can compute the Poisson bracket of the two constraints associated with $N_1$ and $N_2$
\ba
&&\left\{\frac {\pa \mathcal H^{\text{II}}_\text{mini}}{\pa N_1},\,
\frac {\pa \mathcal H^{\text{II}}_\text{mini}}{\pa N_2}\right\}
\nn
&=&
\frac{a_3\,e^{-2B}P_B^2}{8 (a_2 \,e^A + a_3 \,e^B)^4}
\Big[a_1\,e^{A-B}P_B 
\nn&&
\qquad\qquad
-a_3\, e^{-(A-B)}P_A
-a_2\, (P_A-P_B)\Big].
\label{msa-poi}
\ea

If $a_2\neq 0$ and $a_3=0$, the Poisson bracket \eqref{msa-poi} vanishes. 
It seems that the two constraints are first class constraints,  
which could be associated with two sets of gauge symmetries. 
If $a_2\neq 0$ and $a_3\neq 0$, 
an independent constraint  
is generated by the stability of secondary constraints, 
which involves cubic momentum terms according to \eqref{msa-poi}. 
This constraint eliminates one more dynamical variable, 
which signals the complete absence of the sixth degree of freedom. 

Note that it is not justified to take the minisuperspace approximation 
before computing Poisson brackets. 
But this simple computation does capture 
the essential features of 
the Hamiltonian structure: 
\begin{enumerate}
\item
when $a_3=0$, the Poisson brackets of secondary constraints vanish on the constraint surface;
\item
when $a_3\neq 0$, an independent tertiary constraint is generated, 
rendering absent the sixth degree of freedom. 
\end{enumerate}

At this point, it is not clear 
whether the minisuperspace results can be extended to the full theories. 
To verify this, we need to examine the Hamiltonian structure 
of the full theories, 
which is highly technical. 

In the first class of bi-gravity models \eqref{1-class-bigra},  
the kinetic terms are the standard Einstein-Hilbert terms. 
The secondary constraints are 
those in general relativity supplemented 
by the contribution from the potential terms (\ref{bi-gr-pot-2},\,\ref{bi-gr-pot-3},\,\ref{bi-gr-pot-4}), 
which do not involve momenta and spatial derivatives. 
To compute the Poisson brackets of the constraints, 
one can use Dirac's hypersurface deformation algebra. 

However, in the second class of bi-gravity models \eqref{2cl-2}, 
the kinetic part of the Lagrangians are modified 
by the novel kinetic terms. 
To obtain the Hamiltonians already requires some work. 
The constraints have more involved dependence 
on momenta and spatial derivative terms. 
Dirac's algebra is not applicable. 
The computations of constraint brackets are considerably more challenging. 
We leave the technical analysis of the Hamiltonian structure to 
a separate work \cite{Li:2015iwc} 
in which we verify that the sixth degrees of freedom are indeed eliminated,  
and the case $a_2\neq 0,\,a_3=0$ does describe two interacting massless spin-2 fields. 

\section{Field redefinitions}\label{sec-redef}
Let us derive the explicit expressions of the novel kinetic terms. 
It is difficult to achieve this step directly 
because the components of a curvature two-form 
are complicated functions of the associated vielbein. 
To circumvent this difficulty, 
we make use of a mathematical identity for tensors in $d$ dimensions
\footnote{By an abuse of notation, 
the local Lorentz indices are denoted by Greek letters as well. }
\ba
T_{\dots}^{[\m_1\dots\m_d]}
&&=\text{det} (E)\,
T_{\dots}^{\n_1\dots\n_d}
{(E^{-1})_{\n_1}}^{[\m_1}
\dots
{(E^{-1})_{\n_d}}^{\m_d]},\qquad
\label{iden}
\ea
where the antisymmetrized product of $E^{-1}$ gives $\det(E^{-1})$ 
and cancel $\det (E)$  out. 
For example, in two dimensions we have
\ba
&&T^{01}-T^{10}
\nn
&=&\det(E) T^{01}
\big[{(E^{-1})_{0}}^{0}{(E^{-1})_{1}}^{1}-{(E^{-1})_{0}}^{1}{(E^{-1})_{1}}^{0}\big]
\nn&&
+
\det(E) T^{10}
\big[{(E^{-1})_{1}}^{0}{(E^{-1})_{0}}^{1}-{(E^{-1})_{1}}^{1}{(E^{-1})_{0}}^{0}\big].
\nonumber
\ea

We notice the minisuperspace Lagrangian \eqref{mini} contains 
the ratio of two lapse functions. 
From the mathematical identity \eqref{iden},  
it is natural to introduce a new tensor field as the ``ratio'' of two vielbeins
\be
{e_\m}^\n={F_\m}^A{(E^{-1})^\n{}_{A}},\label{field-red}
\ee
where $\m$ and $\n$ are external spacetime indices 
and $A$ is an internal Lorentz index. 

Using the identity \eqref{iden}, the novel kinetic terms become
\ba
\mathcal L_2{}^\text{kin}&=&d^4x \,\sqrt{-g}\,{R(g)_{ab}}^{\m\n}{\d_\m}^{[a}{\d_\n}^{b}{\d_\r}^\r{e_\s}^{\s]},
\nn
\mathcal L_3{}^\text{kin}&=&d^4x \,\sqrt{-g}\,{R(g)_{ab}}^{\m\n}{\d_\m}^{[a}{\d_\n}^{b}{e_\r}^\r{e_\s}^{\s]}.
\ea
After simple manipulations, we have
\ba
\mathcal L_2{}^\text{kin}&=&
(-)\frac 1 4 \,\sqrt{-g}\,{R(g)_{\m\n}}^{[\m\n}{e_\r}^{\r]}\,d^4x,
\label{L2-r}
\\
\mathcal L_3{}^\text{kin}&=&
 \frac 1 4 \sqrt{-g}\,{R(g)_{\m\n}}^{[\m\n}{e_\r}^{\r}{e_\s}^{\s]}\,d^4x,\label{L3-r}
\ea
where the antisymmetrization $[\dots]$ is not normalized. 
The normalization factors in \eqref{L2-r} and \eqref{L3-r} are made precise. 
These choices simplify the expressions below. 
More explicitly, 
$\mathcal L_2{}^\text{kin}$ and $\mathcal L_3{}^\text{kin}$ are functions of the metric $g_{\m\n}$ 
and the new tensor field ${e_\m}^\n$
\ba
\mathcal L_2{}^\text{kin}&=&\sqrt{-g}\,
\left({R_\m}^\n-\frac 1 2 R\, {\d_\m}^\n\right){e_\n}^\m\,d^4x,
\label{L2-R}
\\
\mathcal L_3{}^\text{kin}&=&\sqrt{-g}\,
\Big[
{R_{\m\n}}^{\r\s}{e_\r}^\m {e_\s}^\n
-2{R_\m}^\n
({e_\n}^\m {e_\r}^\r-{e_\n}^\r {e_\r}^\m)
\nn&&\qquad\qquad
+\frac 1 2 R\,
({e_\m}^\m {e_\n}^\n
-{e_\m}^\n {e_\n}^\m)
\Big]\,d^4x.\label{L3-R}
\ea 
In terms of the new tensor field $e_{\m\n}$, the potential terms (\ref{bi-gr-pot-1}-\ref{bi-gr-pot-5})are
\ba
\mathcal L_1{}^\text{pot}&=&\sqrt{-g}\,,\label{L1-p}\\
\mathcal L_2{}^\text{pot}&=&\sqrt{-g}\,e_\m{}^\m,\label{L2-p}\\
\mathcal L_3{}^\text{pot}&=&\sqrt{-g}\,e_\m{}^{[\m} e_\n{}^{\n]},\label{L3-p}\\
\mathcal L_4{}^\text{pot}&=&\sqrt{-g}\,e_\m{}^{[\m} e_\n{}^\n e_\r{}^{\r]}\label{L4-p},\\
\mathcal L_5{}^\text{pot}&=&\sqrt{-g}\,e_\m{}^{[\m} e_\n{}^\n e_\r{}^\r e_\s{}^{\s]}.\label{L5-p}
\ea

We can lower and raise the indices of $e_\m{}^\n$ by the metric $g_{\m\n}$ and its inverse $g^{\m\n}$. For example, we have
\be
e_{\m\n}=e_\m{}^\r g_{\r\n}=F_\m{}^A E_\n{}^B \eta_{AB}.
\ee 
The symmetric condition then translates into
\be
e_{\m\n}=F_\m{}^A E_\n{}^B \eta_{AB}
=F_\n{}^A E_\m{}^B \eta_{AB}=e_{\n\m}
\ee
and we have
\be
e^\m{}_\r\,e^\r{}_\n=g^{\m\a}e_{\a\b}g^{\b\r}e_{\r\n}=g^{\m\a}\,f_{\a\n}. 
\ee
When $e_{\m\n}$ is symmetric, $e^\m{}_\n$ is the square root of $g^{-1}f$ 
with $g_{\m\n},\,f_{\m\n}$ defined in \eqref{metric-def}. 

\section{$\mathcal L_2{}^\text{kin}$ \& quadratic curvature Gravity}\label{sec-L2}
In this section, we focus on the novel kinetic term $\mathcal L_2{}^\text{kin}$ and 
explain its connections to the models of quadratic curvature gravity. 

\subsection{Gauge symmetries of $\mathcal L_2{}^\text{kin}$}
As anticipated in the minisuperspace analysis, 
the case of $a_3=0,\,a_2\neq 0$ seems to be 
related to bi-gravity models with two sets of gauge symmetries. 
Let us emphasis that the coefficient $a_1$ of the Einstein-Hilbert term is not fixed. 

From the explicit form of $\mathcal L_2{}^\text{kin}$, 
we can see linear combination of 
$\mathcal L_1{}^\text{kin}$ and $\mathcal L_2{}^\text{kin}$ 
are invariant under several gauge transformations:
\begin{itemize}
\item
Standard diffeomorphism invariance
\be
\d g_{\m\n}=\pounds_\xi g_{\m\n},\quad
\d e_{\m\n}=\pounds_\xi e_{\m\n}, 
\ee
where $\xi^\m$ is a four-vector. 
This symmetry is expected in a covariant bi-gravity theory. 
\item
Additional ``diffeomorphism invariance"
\be
\d e_{\m\n}=\pounds_{\xi'} g_{\m\n}=\nabla_\m \xi'_\n+\nabla_\n \xi'_\m,
\label{add-diff}
\ee
where $\xi^\m$ is a four-vector. 
The Lagrangians are invariant because $\mathcal L_2{}^\text{kin}$ is 
the product of Einstein tensor and the new tensor $e_{\m\n}$. 
After integrating by parts, the covariant derivative acts on the Einstein tensor 
and the change in the Lagrangians vanish due to the second Bianchi identity. 
\item
Local Lorentz invariance
\be
E_\m{}^A\rightarrow \L_B{}^A\, E_\m{}^B,\quad
F_\m{}^A\rightarrow \L_B{}^A\, F_\m{}^B.
\ee

For an infinitesimal transformation, we have
\be
\d E_\m{}^A= \o_B{}^A\, E_\m{}^B,
\quad
\d F_\m{}^A= \o_B{}^A\, F_\m{}^B,
\ee
where 
\be
\L_B{}^A=\d_B{}^A+\o_B{}^A+\mathcal O(\o^2).
\ee

From the definition \eqref{field-red}, 
we know $e_{\m\n}$ is invariant 
under a diagonal local Lorentz transformation, 
so the Lagrangians are invariant as well.  

\item
Additional ``local Lorentz invariance"
\be
\d e_{\m\n}=t_{\m\n},\quad 
t_{\m\n}=-t_{\n\m}
\ee
or in the infinitesimal form
\be
\d F_\m{}^A=\o'_B{}^A\, E_\m{}^B,\quad \o'_{AB}=-\o'_{BA}. 
\label{add-ll}
\ee

The antisymmetric part of $e_{\m\n}$ is projected out 
by the symmetric Einstein tensor, 
so the Lagrangians are invariant 
under a change in the antisymmetric part of $e_{\m\n}$. 

Since the antisymmetric part drops out when $a_3=0$, 
we can identify the symmetric part of $e^\m{}_\n$ with 
the square root of a metric product $g^{\m\r}f'_{\r\n}$.
\footnote{Note that the metric $f'_{\m\n}$ is different from 
$f_{\m\n}$ defined in \eqref{metric-def} 
and they coincide only when the antisymmetric part of $e_{\m\n}$ vanishes.}

\end{itemize}

These gauge symmetries persist even if we turn on the potential terms 
$\mathcal L_1{}^\text{pot}$ and $\mathcal L_2{}^\text{pot}$, 
which are related to the cosmological constant. 
However, the ``additional" gauge symmetries will be broken when 
$\mathcal L_3{}^\text{kin},\,\mathcal L_3{}^\text{pot},\,\mathcal L_4{}^\text{pot},\,\mathcal L_5{}^\text{pot}$ 
are introduced.

It should be noted that the additional ``invariances" are not the precise transformations according to their names, 
but they do have the same amount and similar forms of gauge symmetries as indicated. 
To be more precise, 
if we substitute $\d e_{\m\n}$ on the left hand side of \eqref{add-diff} with $\d g_{\m\n}$ and 
$\d F_\m{}^A$ on the left hand side of \eqref{add-ll} with $\d E_\m{}^A$, 
they become the off-diagonal gauge transformations. 

\subsection{New Massive Gravity}
One may suspect the bi-gravity models with $a_3=0$ can be transformed into 
two Einstein-Hilbert terms after some field redefinitions. 
It is not clear what field redefinition can make this connection. 
Nevertheless, it was shown in \cite{Paulos:2012xe} that 
$\mathcal L_2{}^\text{kin}$ in \eqref{L2-R} can be obtained by taking a scaling limit of two Einstein-Hilbert terms
\ba
&&\l\left[\sqrt{-f}\,R(f)-\sqrt{-g}\,R(g)
\right]
\nn&\rightarrow 
&\sqrt{-g}\,
\left(R^{\m\n}-\frac 1 2 R\, g^{\m\n}\right)e_{\m\n}
=\mathcal L_2{}^\text{kin}
,
\ea
where $e_{\m\n}$ is defined as 
\be
f_{\m\n}=g_{\m\n}+ e_{\m\n}/\l
\ee
with
\be
\l\rightarrow \infty.
\ee

In fact, $\mathcal L_2{}^\text{kin}$ in the form of \eqref{L2-R} already 
appeared in the auxiliary field representation of 
3d New Massive Gravity \cite{Bergshoeff:2009hq, Bergshoeff:2009aq}, 
which is a theory of quadratic curvature gravity. 

In our language, the Lagrangian of 3d New Massive Gravity reads
\footnote{In this representation, 
we can see one of the BD-ghost-free potentials already appeared in the context of New Massive Gravity. }
\ba
\mathcal L_{NMG}
&=&\s\mathcal L_\text{EH}
+\mathcal L_2{}^\text{kin}
+c_1\mathcal L_1{}^\text{pot}
+c_3 \mathcal L_3{}^\text{pot},
\label{NMG}
\ea
\ba
\mathcal L_\text{EH}&=&R(E)\wedge E, \\
\mathcal L_2{}^\text{kin}&=&R(E)\wedge F, \\
\mathcal L_1{}^\text{pot}&=&E\wedge E\wedge E, \\
\mathcal L_3{}^\text{pot}&=&E\wedge F\wedge F,
\ea
where $\s=\pm 1$ is the sign of the Einstein-Hilbert term, 
$c_1$ is proportional to the cosmological constant, 
$c_3$ corresponds to the mass squared $m^2$.  
The usual auxiliary field is identified with the second tensor field 
\be
e_{\m\n}=e_\m{}^\r g_{\r\n}=F_\m{}^A E_\n{}^B \eta_{AB}.
\ee  

Note that the symmetric condition 
\be
e_{\m\n}=e_{\n\m}
\ee
is imposed dynamically by the equations of motion
\footnote{In section \ref{sec-sym}, 
we discuss how to generalize this example 
of dynamical symmetric condition.}. 

Therefore, New Massive Gravity is an example of 
3d bi-gravity models in the second class. 
It is known that 3d New Massive Gravity do not contain the Boulware-Deser ghost \cite{ham-NMG}, 
which furnishes evidence that the second class of bi-gravity models 
are free of the BD ghost.  

A straightforward generalization of 3d New Massive Gravity is to introduce 
other potential terms $\mathcal L_2{}^\text{pot},\,\mathcal L_4{}^\text{pot}$ \cite{Paulos:2012xe}
\ba
\mathcal L_2{}^\text{pot}&=&E\wedge E\wedge F, \\
\mathcal L_4{}^\text{pot}&=&F\wedge F\wedge F.
\ea 
Note that, if $\mathcal L_4{}^\text{pot}$ is considered, 
the Lagrangian does not reduce to that of quadratic curvature gravity 
when $e_{\m\n}$ is integrated out. 
Instead, it contains infinitely many higher order curvature corrections. 

In 3d New Massive Gravity, the Einstein-Hilbert term is always present. 
One of the reasons may be that 
the second spin-2 field $e_{\m\n}$ is usually 
considered to be an auxiliary field, 
which seems to have no dynamics. 
We want to emphasize that $\mathcal L_2{}^\text{kin}$ is a kinetic term as well, 
and the use of the Einstein-Hilbert term is not necessary
\footnote{To eliminate the second order time-derivative terms due to the curvature tensor in $\mathcal L_2{}^\text{kin}$, 
we need to supplement the action by boundary terms analogous to the York-Gibbons-Hawking term. 
Then $\mathcal L_2{}^\text{kin}$ generates a time derivative term $\pa_t e_{\m\n}$ in the Lagrangian, 
so both $g_{\m\n}$ and $e_{\m\n}$ have dynamical degrees of freedom. 
Furthermore, if we expand the Lagrangian around a Minkowski background 
and diagonalize the quadratic kinetic terms, 
$\mathcal L_2{}^\text{kin}$ will give rise to two linearized Einstein-Hilbert terms. }.
Therefore, the Einstein-Hilbert term could be absent, 
then the 3d bi-gravity Lagrangian reads 
\be
\mathcal L=a_2\, \mathcal L_2{}^\text{kin}+\sum_{i=1}^4 c_i\,\mathcal L_i{}^\text{pot},
\quad a_2\neq 0.\label{3d-no-EH}
\ee  
The number of dynamical degrees of freedom should be 
the same as that of New Massive Gravity 
and the Boulware-Deser ghost should not be not propagating. 
The cases without the Einstein-Hilbert term are related to 
the generalized NMG in \cite{Paulos:2012xe} 
by a field redefinition
\be
e_{\m\n}\rightarrow e_{\m\n}+ c\, g_{\m\n}. 
\ee
where $c$ depends on the coefficients of $\mathcal L_\text{EH}$ and $\mathcal L_2{}^\text{kin}$. 

In 3d, 
$\mathcal L_2{}^\text{kin}$ is the only novel kinetic term from \eqref{multi-kin} 
due to the limited number of spacetime indices. 
Since a massless graviton in 3d has no dynamical degree of freedom, 
we can choose $a_2$ such that 
the kinetic term of the massive graviton has a correct sign. 
Then \eqref{3d-no-EH} is a unitary theory of 3d massive gravity. 
The special case of $c_1=c_2=c_4=0$ was discussed in \cite{Deser:2009hb} 
and that of $c_1\neq 0, \, c_2=c_4=0$ in \cite{Bergshoeff:2011ri}.

\subsection{Critical gravity}
There exists a continuous family of critical points \cite{Liu:2009bk}
in the parameter space of 
3d New Massive Gravity $\eqref{NMG}$ 
and its higher dimensional generalization \cite{critical-gr} , 
\ba
\mathcal L&=& R(E)\wedge E \wedge\dots\wedge (E+F)
\nn&&+
 E \wedge\dots\wedge E
\wedge(\L\,E \wedge E+m^2\,F\wedge F),
\ea
which are known as critical gravity.  
At these critical points, the cosmological constant $\L$ is proportional to 
the mass squared $m^2$ with dimension-dependent coefficients. 
Integrating out the auxiliary field, 
the linearized 4th order equation of motions contains two massless spin-2 modes
\footnote{However, the total number of dynamical degrees of freedom 
should be the same as 
that of 1 massless and 1 massive gravitons. 
The second massless graviton seems to be an artifact of 
the linearized equation of motion at the critical points, 
as there is no symmetry enhancement. 
For example, logarithmic modes are allowed 
if we do not assume the Brown-Henneaux boundary conditions 
\cite{Brown:1986nw}. 
They were claimed to be dual to Logarithmic Conformal Field Theories 
\cite{Grumiller:2008qz, Grumiller:2009sn, Alishahiha:2011yb, Bergshoeff:2011ri}.}. 

Here we want to point out that 
\be
\mathcal L_2{}^\text{kin}
=R(E)\wedge F
\ee
and its higher dimensional version
\be
\mathcal L_2{}^\text{kin}
=R(E)\wedge E\wedge \dots \wedge E \wedge F
\label{1-F-kin}
\ee
can be thought of as 
a special limit of critical gravity in the bi-gravity formulation
\footnote{The Einstein-Hilbert term is absorbed into $\mathcal L_2{}^\text{kin}$ 
by redefining $F$.}. 
Note that in this limit, $e_{\m\n}$ can not be integrated out 
because it is a Lagrange multiplier.

It is shown in the Hamiltonian analysis of \cite{Li:2015iwc} 
that $\mathcal L_2{}^\text{kin}$ in 4d has 
two sets of first-class constraints, 
corresponding to two sets of gauge symmetries. 
More general, the Lagrangian
\be
\mathcal L=a_1\,\mathcal L_1{}^\text{kin}
+a_2\,\mathcal L_2{}^\text{kin}
+c_1\,\mathcal L_1{}^\text{pot}
+c_2\,\mathcal L_2{}^\text{pot}
\ee
with
\be
a_2\neq 0
\ee 
are interacting theories of two massless gravitons in various dimensions ($d>2$), 
where $c_1$ and $c_2$ are related to the cosmological constant
\footnote{There is a no-go theorem for interacting theories 
of massless, gauge invariant, spin-2 fields 
if the Lagrangian has at most two derivatives 
\cite{Boulanger:2000rq}. 
This is not in contradiction to the present work, 
because one of the linearized kinetic terms has a wrong sign, 
which violates one of the assumptions in the no-go theorem. 
The details of the linearized actions are discussed in section \ref{sec-lin}. 
A recent construction of color-decorated gravity \cite{Gwak:2015vfb} 
evades this no-go theorem by including extra fields. }.  

The same gauge symmetries are also realized in higher derivative counterparts 
of the two-derivative term $\mathcal L_2{}^\text{kin}$ 
\be
R(E)\wedge\dots \wedge R(E)\wedge E\wedge \dots \wedge E \wedge F,
\label{bi-Lovelock}
\ee
where one of the E-vielbeins in the Lovelock terms \cite{Lovelock:1971yv} 
is replaced by an F-vielbein. 
The additional symmetries are due to the fact that 
Lovelock tensors are both symmetric and divergence-free. 

Along the line of the second class bi-gravity theories \eqref{2cl-2}, 
we propose a general Lagrangian describing 
two interacting, massless, gauge invariant gravitons, 
which is a linear combination of Lovelock terms, 
the novel derivative terms (\ref{1-F-kin}, \ref{bi-Lovelock})
and two potential terms 
\be
\mathcal L_1{}^\text{pot}=E\wedge\dots \wedge E,
\quad\mathcal L_2{}^\text{pot}=E\wedge\dots \wedge E\wedge F,
\ee 
where at least one of the novel derivative terms is present 
\footnote{Deforming these massless models 
by other potential terms with two vielbeins, 
one obtains the generalizations of New Massive Gravity proposed in \cite{Paulos:2012xe}.}.

\subsection{Weyl gravity}
Weyl gravity is a well-known theory of quadratic curvature gravity in 4d, 
which is both diffeomorphism and conformal invariant. 
Interestingly, the Lagrangian of Weyl gravity in the two-derivative representation \cite{Kaku:1977pa} 
has a compact form in terms of Vielbeins 
\footnote{By redefining $F=F'+E$, one have the auxiliary field reformulation of Weyl gravity with an Einstein-Hilbert term and a cosmological constant term. 
where the second spin-2 field $e_{\m\n}$ can be thought of as a matter field couples to the Einstein tensor. }
\be
\mathcal L_\text{Weyl}=R(E)\wedge E\wedge F+ E\wedge E \wedge F \wedge F, 
\label{Weyl}
\ee
where $F$ has dimension 2. 

Then the absence of Ostrogradsky's scalar ghost is translated into 
the absence of the BD ghost. 
Ostrogradsky's spin-2 ghost in the four-derivative formulation 
now becomes a basic spin-2 ghost due to a wrong sign kinetic term. 

In this representation, 
Weyl gravity is built from a novel kinetic term and a dRGT potential term.
Despite the presence of a spin-2 ghost, Weyl gravity is the first example of 
nonlinear completions of Fierz-Pauli massive gravity that are free of the BD ghost, 
where the nonlinear theory was proposed 20 years before the linear one. 

Furthermore, Weyl gravity is a special bi-gravity model in the second class  \eqref{2cl-2}
with an emergent gauge symmetry (conformal symmetry). 
This gauge symmetry is a nonlinear completion of 
the additional gauge symmetry 
of a massive spin-2 field around de-Sitter background 
at the partially massless point \cite{Deser:2012qg}.
To make this connection more clear, 
we linearize \eqref{Weyl} around de-Sitter background and 
diagonalize the quadratic Lagrangian in the next section.

\section{Linearized Lagrangians}\label{sec-lin}
In the previous section, 
we show that some of the bi-gravity models with novel kinetic terms are equivalent to higher derivative gravity models. 
It is well known that higher derivative gravity models usually contain spin-2 ghosts, 
which could lead to the problem of non-unitarity. 
In this section, we derive the quadratic actions of 
the novel kinetic terms and examine  
whether this is a general feature of 
the bi-gravity theories in the second class \eqref{2cl-2}. 

\subsection{Minkowski background}
Consider a bi-gravity model in 4d whose Lagrangian reads
\be
\mathcal L=a_1\,\mathcal L_\text{EH}
+a_2\, \mathcal L_2{}^\text{kin}
+a_3\, \mathcal L_3{}^\text{kin},
\ee
where $\mathcal L_\text{EH}$ is the Einstein-Hilbert kinetic term 
\be
\mathcal L_\text{EH}=\sqrt{-g}\, R(g),
\ee
$\mathcal L_2{}^\text{kin}$ and 
$\mathcal L_3{}^\text{kin}$ are 
the novel nonlinear kinetic terms 
defined in (\ref{L2-r}, \ref{L3-r}).

Let us expand the metric field $g_{\m\n}$ 
and the symmetric tensor field $e_{\m\n}$ 
around the Minkowski background
\be
g_{\m\n}=\eta_{\m\n}+\d g_{\m\n},
\ee
\be
e_{\m\n}=\eta_{\m\n}+\d e_{\m\n}, 
\ee
where we assume there is no numerical factors in front of $\eta_{\m\n}$. 
These factors can always be absorbed into $a_1,\,a_2,\,a_3$ by redefining 
$g_{\m\n}$ and $e_{\m\n}$. 

Note that $e_{\m\n}$ simply vanishes 
in a different kind of background solutions. 
The two kinds of background solutions are related 
by a redefinition of $F_\m{}^A$
\be
F'=F+E.
\ee 

To the quadratic order, the linearized Lagrangian reads
\be
\bar{\mathcal L}=
c_1\, \d g_\m{}^{[\m}\pa_\n\pa^\n \d g_\r{}^{\r]}
+c_2\, \d g_\m{}^{[\m}\pa_\n\pa^\n \d e_\r{}^{\r]},
\label{lin-b}
\ee
where the first term is the linearized Einstein-Hilbert term 
and the coefficients are
\be
c_1=\frac 1 4(- a_1+a_2- a_3),\quad
c_2=\frac 1 2a_2-a_3.
\ee

Now we can diagonalize the quadratic Lagrangian
\be
\bar{\mathcal L}=c_1\left( h_\m{}^{[\m}\pa_\n\pa^\n h_\r{}^{\r]}
-H_\m{}^{[\m}\pa_\n\pa^\n H_\r{}^{\r]}\right),
\ee
where we assume $c_1\neq 0$ and the diagonalized spin-2 fields are
\be
h_{\m\n}=\d g_{\m\n}+\frac {c_2}{2c_1}\d e_{\m\n}, 
\quad H_{\m\n}=\frac{c_2}{2c_1}\d e_{\m\n}.
\ee

If $c_1=0$ and $c_2\neq 0$, 
then the first term in \eqref{lin-b} vanishes 
and the diagonalized Lagrangian is
\be
\bar{\mathcal L}=c_2\left( h_\m{}^{[\m}\pa_\n\pa^\n h_\r{}^{\r]}
-H_\m{}^{[\m}\pa_\n\pa^\n H_\r{}^{\r]}\right),
\ee
and the diagonalized fields are
\be
h_{\m\n}=\frac 1 2(\d g_{\m\n}+\d e_{\m\n}), 
\quad H_{\m\n}=\frac 1 2(\d g_{\m\n}-\d e_{\m\n}).
\ee

The linearized Lagrangian 
after diagonalization is a linear combination of 
two linearized Einstein-Hilbert terms 
\footnote{The diagonalized form is invariant under 
a continuous family of field redefinitions
\ba
h_{\m\n}=\cosh(\theta) h'_{\m\n}+\sinh(\theta)H'_{\m\n},\nn
H_{\m\n}=\sinh(\theta) h'_{\m\n}+\cosh(\theta)H'_{\m\n}.
\ea} 
except in some special cases. 
The diagonalized kinetic terms always have opposite signs 
due to the absence of quadratic term of $e_{\m\n}$, 
which can be traced back to the absence of $R(F)$. 

The special cases are
\be
c_2=0,\quad \text{or}\quad \frac 1 2 a_2=a_3,
\ee
then $H_{\m\n}=0$ and 
the second diagonalized kinetic term vanishes. 
The first kinetic term has a right sign when 
\be
c_1=\frac 1 4(- a_1+ a_3)<0.
\ee

A more extreme case is 
\be
c_1=c_2=0,\quad
\text{or}\quad
a_1=\frac 1 2 a_2=a_3,
\ee
then the linearized Lagrangian is empty. 
In these special cases, 
the bi-gravity models are strongly coupled due to 
the absence of some linearized kinetic terms. 

Note that the Lagrangians of these special cases can be 
schematically written as
\be
\mathcal L=(a_1-a_3)\,R\wedge E\wedge E+a_3\, R\wedge (F-E)\wedge (F-E).
\ee
If we consider the other kind of background solutions where $\bar e_{\m\n}$ vanishes, 
both of the two linearized kinetic terms are present and 
there is no issue of strong coupling. 
So the problem of strong coupling depends on 
the choice of background solutions. 
\footnote{If we consider backgrounds that $e_{\m\n}$ vanish, 
then $\mathcal L_3{}^\text{kin}$ will not contribute to the linearized Lagrangian. 
We will encounter the strong coupling problem when $a_2=0$.}
The existence of these strongly coupled backgrounds is related to 
the presence of spin-2 ghosts. 

The potential terms (\ref{L1-p}-\ref{L5-p}) can generate linear terms around a Minkowski background, 
which signals a wrong choice of vacuum. 
In the next subsection, we discuss the linearized actions 
around general maximally symmetric backgrounds.

\subsection{Constant curvature background}
Let us introduce nonlinear potential terms to the 4d Lagrangian
\ba
\mathcal L&=&a_1\,\mathcal L_\text{EH}
+a_2\,\mathcal L_2{}^\text{kin}
+a_3 \,\mathcal L_3{}^\text{kin}
\nn&&
+b_1\,\mathcal L_1{}^\text{pot}
+b_2\,\mathcal L_2{}^\text{pot}
+b_3 \,\mathcal L_3{}^\text{pot}
\nn&&
+b_4\,\mathcal L_4{}^\text{pot}
+b_5 \,\mathcal L_5{}^\text{pot},
\label{lin-pot}
\ea
where the potential terms are defined in 
\eqref{L1-p}-\eqref{L5-p}.

The spin-2 fields $g_{\m\n}$ and $e_{\m\n}$ 
are expanded around a cosmological background $\bar g_{\m\n}$
\ba
g_{\m\n}&=&\bar g_{\m\n}+\d g_{\m\n},
\nn
e_{\m\n}&=&\d e_{\m\n},
\ea
where the background spacetime has constant curvature
\be
\bar R_{\m\n\r\s}=\frac  \L 3
(\bar g_{\m\r}\,\bar g_{\n\s}
-\bar g_{\m\s}\,\bar g_{\n\r}).
\ee

We assume the background value of $e_{\m\n}$ vanishes, 
which is not necessary. 
However, if the background solution of $e_{\m\n}$ is proportional to the background metric $\bar g_{\m\n}$, 
then we can always set it to zero by a shift in $F_\m{}^A$
\footnote{The definitions of fluctuating fields are modified accordingly.}. 
 
To avoid strong coupling problem, 
we require 
\be
a_2\neq 0, 
\ee
otherwise we should consider a different background solution. 
 
Around a background solution, 
the linear terms in the perturbative Lagrangian should vanish, 
which indicates
\be
b_1=-2a_1\,\L,\quad b_2=a_2\,\L. 
\ee
Before the shift in $F$, the two equations  
correspond to the solution of the background metric $\bar g_{\m\n}$ 
and the ratio between two background spin-2 fields. 
 
The linearized Lagrangian is
\ba
\bar{\mathcal L}&=&
\sqrt{-\bar g}\,c_1 \left(
\d g_\m{}^{[\m}\bar\nabla_\n\bar\nabla^\n\d g_\r{}^{\r]}
+ \d g_\m{}^{\n}[\bar\nabla_\r,\,\bar\nabla^\m]\d g_\n{}^{\r}\right)
\nn&&
+\sqrt{-\bar g}\,c_2\left(
\d e_\m{}^{[\m}\bar\nabla_\n\bar\nabla^\n\d g_\r{}^{\r]}
+ \d e_\m{}^{\n}[\bar\nabla_\r,\,\bar\nabla^\m]\d g_\n{}^{\r}\right)
\nn&&
+\sqrt{-\bar g}\,\Big[2a_1\L 
-a_1\L\Big(\frac 1 4 \d g_\m{}^\m\d g_\n{}^\n
-\frac 1 2 \d g_\m{}^\n\d g_\n{}^\m\Big)
\nn&&\qquad\qquad
+a_2\,\L\Big(\frac 1 2 \d g_\m{}^\m \d e_\n{}^\n
-\d g_\m{}^\n \d e_\n{}^\m\Big)
\nn&&\qquad\qquad
+b_3\, (\d e_\m{}^\m\d e_\n{}^\n
-\d e_\m{}^\n\d e_\n{}^\m)
\Big]
\ea
where the coefficients 
\be
c_1=-\frac 1 4 a_1,
\quad
c_2=\frac 1 2 a_2
\ee
are simplified due to a shift in $F$. 
Total derivative terms are neglected. 

If $a_1\neq 0$, the diagonalized fields are
\be
h_{\m\n}=\d g_{\m\n}-\frac {a_2}{a_1} \d e_{\m\n},
\quad
H_{\m\n}=\frac {a_2}{a_1} \d e_{\m\n},
\ee
the linearized Lagrangian becomes
\ba
\bar{\mathcal L}&=&
-\frac 1 4 a_1\sqrt{-\bar g}\, \left(
h_\m{}^{[\m}\bar\nabla_\n\bar\nabla^\n h_\r{}^{\r]}
+ h_\m{}^{\n}[\bar\nabla_\r,\,\bar\nabla^\m]h_\n{}^{\r}\right)
\nn&&
+\frac 1 4 a_1\sqrt{-\bar g}\,\left(
H_\m{}^{[\m}\bar\nabla_\n\bar\nabla^\n H_\r{}^{\r]}
+H_\m{}^{\n}[\bar\nabla_\r,\,\bar\nabla^\m] H_\n{}^{\r}\right)
\nn&&
+\sqrt{-\bar g}\,\Big[
-a_1\L\Big(\frac 1 4 h_\m{}^\m h_\n{}^\n
-\frac 1 2 h_\m{}^\n h_\n{}^\m\Big)
\nn&&\qquad\qquad
+a_1\L\Big(\frac 1 4 H_\m{}^\m H_\n{}^\n
-\frac 1 2 H_\m{}^\n H_\n{}^\m\Big)
\nn&&\qquad\qquad
+b_3\Big(\frac {a_1}{a_2}\Big)^2\, (H_\m{}^\m H_\n{}^\n
-H_\m{}^\n H_\n{}^\m)
\Big]
\ea
where the constant term is neglected. 
The first four lines are the linearized Einstein-Hilbert terms 
with cosmological constant $\L$ around the background solutions. 
The last line is the Fierz-Pauli mass term for $H_{\m\n}$. 
The coefficient of the massless spin-2 field $h_{\m\n}$ is $a_1$, 
while that of $H_{\m\n}$ is $-a_1$. 
One of them is a spin-2 ghost. 
The mass squared of $H_{\m\n}$ is determined by $b_3$. 

If $a_1=0$, the diagonalized fields are
\be
h_{\m\n}=\frac 1 2(\d g_{\m\n}+\d e_{\m\n}), 
\quad H_{\m\n}=\frac 1 2(\d g_{\m\n}-\d e_{\m\n}).
\ee
In addition, $b_3$ should vanish in order to be consistent with our choice of background solution $\bar e_{\m\n}=0$,
so the mass terms vanish. 
The linearized Lagrangian is
\ba
\bar{\mathcal L}&=&
\frac 1 2 a_2\sqrt{-\bar g}\, \left(
h_\m{}^{[\m}\bar\nabla_\n\bar\nabla^\n h_\r{}^{\r]}
+ h_\m{}^{\n}[\bar\nabla_\r,\,\bar\nabla^\m]h_\n{}^{\r}\right)
\nn&&
-\frac 1 2 a_2\sqrt{-\bar g}\,\left(
H_\m{}^{[\m}\bar\nabla_\n\bar\nabla^\n H_\r{}^{\r]}
+H_\m{}^{\n}[\bar\nabla_\r,\,\bar\nabla^\m] H_\n{}^{\r}\right)
\nn&&
+\sqrt{-\bar g}\,\Big[
2a_2\L\Big(\frac 1 4 h_\m{}^\m h_\n{}^\n
-\frac 1 2 h_\m{}^\n h_\n{}^\m\Big)
\nn&&\qquad\qquad
-2a_2\L\Big(\frac 1 4 H_\m{}^\m H_\n{}^\n
-\frac 1 2 H_\m{}^\n H_\n{}^\m\Big)
\Big],
\ea
which corresponds to two interacting massless gravitons. 

\subsection{Linearized Weyl gravity}
In this subsection, we would like to discuss 
the linearized action of Weyl gravity \eqref{Weyl} 
around the de-Sitter background. 
As shown in \cite{Deser:2012qg} , 
the conformal transformation in Weyl gravity can be recast into 
nonlinear partially massless transformation for spin-2 matter field 
after some field redefinitions. 
So we expect after diagonalization the massive spin-2 field 
has linear partially massless gauge symmetry. 

The explicit expression of \eqref{Weyl} is
\ba
\mathcal L_\text{Weyl}&=&
\mathcal L_2{}^\text{kin}
+\mathcal L_3{}^\text{pot}
\nn&=&
\sqrt{-g}\Big[\Big(R^{\m\n}-\frac 1 2  R\,g^{\m\n}\Big)e_{\m\n}
\nn&&\qquad\quad
+e_\m{}^\m\, e_\n{}^\n-e_{\m\n}\,e^{\m\n}\Big], 
\ea
where the equations of motion for $e_{\m\n}$ is
\be
e_{\m\n}=\frac 1 2 R_{\m\n}-\frac 1 {12} R\, g_{\m\n}.
\ee

The nonlinear gauge symmetry transformations are
\begin{itemize}
\item
conformal invariance
\be
g_{\m\n}\rightarrow (1+2\phi) g_{\m\n},\quad
e_{\m\n}\rightarrow e_{\m\n}-\nabla_\m\pa_\n\phi;
\ee

\item
diffeomorphism invariance
\be
g_{\m\n}\rightarrow g_{\m\n}+\pounds_\xi g_{\m\n},\quad
e_{\m\n}\rightarrow e_{\m\n}+\pounds_\xi e_{\m\n}.
\ee

\end{itemize}

Let us consider the de-Sitter background solution
\be
\bar g_{\m\n}=g_{\m\n}^\text{dS},\quad
\bar e_{\m\n}=\frac \L 6 g_{\m\n}^\text{dS}, 
\ee
where we keep the non-zero background value of $e_{\m\n}$. 

The fluctuations around the de-Sitter vacuum are
\be
\d g_{\m\n}=g_{\m\n}-\bar g_{\m\n},\quad
\d e_{\m\n}=e_{\m\n}-\bar e_{\m\n}. 
\ee

Then we can expand the full action to the quadratic order
\ba
\bar{\mathcal L}_\text{Weyl}&=&
\sqrt{-\bar g}\Big[-\frac {\L^2}3+(\d e^2-\d e_{\m\n}\d e^{\m\n})
\nn&&
+\frac \L 6(\d e \d g-4\d e_{\m\n}\d g^{\m\n})
-\frac {\L^2} {72}(\d g^2-4\d g_{\m\n}\d g^{\m\n})
\Big]
\nn&&
+\frac 1 2\sqrt{-\bar g}\left(
\d e_\m{}^{[\m}\bar\nabla_\n\bar\nabla^\n\d g_\r{}^{\r]}
+ \d e_\m{}^{\n}[\bar\nabla_\r,\,\bar\nabla^\m]\d g_\n{}^{\r}\right)
\nn&&
-\frac \L {24}\sqrt{-\bar g}\left(
\d g_\m{}^{[\m}\bar\nabla_\n\bar\nabla^\n\d g_\r{}^{\r]}
+ \d g_\m{}^{\n}[\bar\nabla_\r,\,\bar\nabla^\m]\d g_\n{}^{\r}\right),
\nn
\ea
where total derivative terms are neglected, 
the covariant derivative $\bar \nabla$ is compatible with 
the background metric $\bar g_{\m\n}$ and 
\be
\d g=\d g_\m{}^\m,\quad \d e=\d e_\m{}^\m.
\ee

The diagonalized fields are
\be
h_{\m\n}=\frac {6}\L \d e_{\m\n},\quad
H_{\m\n}=\d g_{\m\n}-\frac {6}\L \d e_{\m\n}.
\ee

The quadratic Lagrangian in terms of $h_{\m\n},\,H_{\m\n}$ reads
\ba
&&\bar{\mathcal L}_\text{Weyl}
\nn
&=&
\frac \L {24}\sqrt{-\bar g}\left(
h_\m{}^{[\m}\bar\nabla_\n\bar\nabla^\n h_\r{}^{\r]}
+ h_\m{}^{\n}[\bar\nabla_\r,\,\bar\nabla^\m]h_\n{}^{\r}\right)
\nn&&
-\frac{\L^2}{48}\sqrt{-\bar g}\left(h_{\m\n}h^{\m\n}-\frac 1 2h^2\right)
\nn&&
-\frac \L {24}\sqrt{-\bar g}\left(
H_\m{}^{[\m}\bar\nabla_\n\bar\nabla^\n H_\r{}^{\r]}
+ H_\m{}^{\n}[\bar\nabla_\r,\,\bar\nabla^\m]H_\n{}^{\r}\right)
\nn&&
+\frac {\L^2}{18}\sqrt{-\bar g}\left(H_{\m\n}H^{\m\n}-\frac 1 4H^2\right), 
\ea
where we neglect the constant term. 
We can see $h$ is a massless spin-2 field with a negative Planck mass, 
while $H$ is a massive spin-2 field with a positive Planck mass. 
The signs of the kinetic terms are  opposite. 

We can further examine the gauge symmetries at the linearized level:
\begin{itemize}
\item
linearized conformal symmetry
\ba
\d g_{\m\n}&\rightarrow& \d g_{\m\n}+2\phi \bar g_{\m\n},\\
\d e_{\m\n}&\rightarrow& \d e_{\m\n}-\bar\nabla_\m\pa_\n\phi,
\ea
and
\ba
h_{\m\n}&\rightarrow&h_{\m\n}-\frac {6}\L \bar\nabla_\m\pa_\n\phi,\\
H_{\m\n}&\rightarrow&H_{\m\n}+\left( \bar\nabla_\m\pa_\n+\frac \L 3\bar g_{\m\n}\right)\left(\frac {6}\L\phi\right).
\ea

\item 
linearized diffeomorphism symmetry
\ba
\d g_{\m\n}&\rightarrow& \d g_{\m\n}+\bar\nabla_\m \xi_\n+\bar\nabla_\n \xi_\m,\\
\d e_{\m\n}&\rightarrow& \d e_{\m\n}+\frac \L 6\left(\bar\nabla_\m \xi_\n+\bar\nabla_\n \xi_\m\right),
\ea
and
\ba
h_{\m\n}&\rightarrow& h_{\m\n} +\bar\nabla_\m \xi_\n+\bar\nabla_\n \xi_\m,\\
H_{\m\n}&\rightarrow& H_{\m\n}.
\ea
\end{itemize}

Therefore, the linear partially massless gauge transformation
\be
H_{\m\n}\rightarrow
H_{\m\n}+\left( \bar\nabla_\m\pa_\n+\frac \L 3\bar g_{\m\n}\right)\a
\ee
is a combination of linearized conformal and diffeomorphism transformations. 

Interestingly, 
only the massless spin-2 field $h$ transforms 
under a change of coordinate. 
The massive spin-2 field can not transform 
because the Lagrangian of massive mode is not invariant.

In this way, we provide a different perspective of Weyl gravity 
by using a novel kinetic term and a dRGT term. 
In this representation, one may understand better 
why unitary partially massless gravity in 4d 
is not found \cite{Joung:2014aba}
\footnote{To be more precise, we require the off-shell action to be gauge invariant.}. 
Along the line of dRGT massive gravity, 
there have been many recent investigations on nonlinear partially massless gravity \cite{deRham:2012kf, deRham:2013wv, PMG}. 
In 4d,  a promising candidate was identified in dRGT massive gravity
\cite{deRham:2012kf} , 
which makes use of precisely the same potential term
\be
E\wedge E \wedge F \wedge F, 
\ee
but the kinetic term is assumed to be the Einstein-Hilbert term and 
the $F$ vielbein is fixed to be de-Sitter. 
Partially massless gauge symmetry is only an artifact 
of the perturbative Lagrangian at low orders. 
We also confirm the suspicion in \cite{deRham:2013wv} that a new kinetic term is required 
in order to extend the partially massless symmetry to the nonlinear level, 
which becomes trivial from our bi-gravity reformulation of Weyl gravity. 

The fact that the novel kinetic terms have no non-trivial single dynamical metric limit 
indicates that we can not truncate Weyl gravity in a consistent manner 
to obtain a nonlinear theory of partially massless gravity 
with single dynamical metric and a fixed de-Sitter metric.  

In addition, partially massless gauge symmetry in 4d can be thought of as an emergent gauge symmetry 
of the 4d bi-gravity models constructed from novel kinetic terms and dRGT potential terms. 
It is tempting to consider the case of three dimensions, 
where a massless spin-2 field has no dynamical degree of freedom 
and we do not need to worry about the sign of its kinetic term. 
However, there is only one new kinetic term in 3d, 
and these bi-gravity models were well investigated 
along the line of 3d New Massive Gravity. 
In particular, the 3d version of Weyl gravity proposed in \cite{Deser:2009hb} 
is an example of the conflict between diffeomorphism and conformal invariances in three dimensions. 
Partially massless symmetry has no nonlinear extension in 3d up to date.

\section{Spin-2 ghost}\label{sec-s2gh}
In the previous section, 
we show that, generically, the two linearized kinetic terms have opposite signs. 
A kinetic term with a wrong sign usually results in 
an unbounded Hamiltonian and non-unitarity. 
If we associate the Hamiltonian with the energy, 
then we could extract infinite energy from a system 
whose Hamiltonian is unbounded from below, 
which leads to classical instability. 
Non-unitarity in a quantum theory means unphysical negative probability. 
Therefore, when a kinetic term has a wrong sign, 
the corresponding degree of freedom is considered to be an unwanted ``ghost". 
\\

From the classical point of view, 
it is not clear which is the correct sign for a massless spin-2 kinetic term, 
as the Hamiltonian simply vanishes on-shell. 
Naively, the Einstein-Hilbert term seems to have a wrong sign 
in the minisuperspace approximation \eqref{mss-1}
\be
\sqrt{-g}\,R(g)\rightarrow \,- \,\frac {12} N (\dot A)^2 e^{3A}, \quad N>0,
\ee
but the Hamiltonian is still bounded because it vanishes. 
If we modify the sign of the Einstein-Hilbert term, 
Newton's constant will become negative and 
gravity be a repulsive force. 
Certainly, this contradicts with our physical world, 
but this is not ruled out as a theoretical possibility. 
We know the Coulomb force is repulsive for like charges. 

The Hamiltonian of novel kinetic terms vanishes on-shell as well \cite{Li:2015iwc}, 
which is expected in covariant theories, 
so a bounded Hamiltonian is not a strong evidence for classical stability. 
One might need to examine other definitions of energy. 
As the local definition of gravitational energy is controversial, 
it may be more sensible to consider global energies (masses) 
according to the isometries of asymptotic spacetime. 
They are the conserved charges associated with the global symmetries of the vacuum 
where the infinite-dimensional diffeomorphism group is spontaneously broken to 
a finite-dimensional global symmetry group. 
In critical gravity models, 
the Abbott-Deser-Tekin mass \cite{ADT} of black hole solutions 
were shown to be zero \cite{critical-gr}. 
As discussed before, the novel kinetic term $\mathcal L_2{}^\text{kin}$ 
can be considered as the special limit of critical gravity, 
and we expect it has the same property. 
\\

From the quantum perspective, 
the correct sign for spin-2 kinetic terms has a more definite answer. 
Unitarity requires particle poles to have positive residues. 
A propagator can be derived from the quadratic action, 
so the correct signs of the kinetic term and the mass term are determined. 

The residue of a spin-2 propagator depends on the coupling to matter. 
It vanishes if the spin-2 field is not coupled to the energy-momentum tensor. 
So one can avoid negative residues 
by identifying the physical metric with 
the spin-2 field with a correct sign kinetic term. 
Using the effective metric \cite{eff-metric}, 
we can escape the problem of tree-level non-unitarity.
\\

Furthermore, 
to obtain the solutions of a model we need boundary conditions. 
When the Lagrangian allows for ghost-like excitations, 
they can still be avoided by proper boundary conditions. 
In this way, we could eliminate the ghost-like modes 
whose kinetic terms have wrong signs, 
then the bi-gravity models should reduce to 
healthy vector-tensor theories 
when the mass squared has a correct sign. 

\section{Cut-off scale}\label{sec-cutoff}
Ghosts are ubiquitous in the framework of effective field theories. 
Their presence do not disqualify the models from describing nature. 
They just tell us when the theories stop providing consistent descriptions 
and microscopic details become important.  

As an effective field theory, 
Einstein's gravity has a cut-off scale 
set by the Planck mass. 
Higher curvature terms are also compatible with 
diffeomorphism invariance, 
so they should be present. 
The natural values of their coefficients are of order unity in terms of the Planck mass, 
and the cut-off scale remains the same. 
Ostrogradsky's ghosts due to higher derivative equations of motion 
are not excited below the Planck scale 
because their masses are around the Planck value. 
The corrections due to higher curvature terms are negligible at low energy scale. 

However, if for some unknown reasons, 
the coefficients of some correction terms are considerably larger than their natural values, 
then we need to worry about Ostrogradsky's instability even below the Planck scale. 
The cut-off scale of an effective field theory of gravity is lowered by the ghost modes. 

Let us consider an example in 4d which admits a bi-gravity reformulation 
\ba
\mathcal L&=&M_p^2\sqrt{-g}\,
\big[R+R\,( e+e\,e+\dots)
\nn&&\qquad+ \L(1+e)+m^2\,( e\, e+e\, e\, e+\dots)\big],
\label{bigra-gen}
\ea
where the tensor structures are not written explicitly, 
the two spin-2 fields $g_{\m\n},\,e_{\m\n}$ are dimensionless. 
After linearization and diagonalization, 
$m^2$ corresponds to the mass of the massive spin-2 field which is ghost-like
\footnote{It is assumed that the coefficients of the potential terms are of the same order, 
which is not a necessary assumption. 
When their magnitude are different, the smallest one is the most important. }. 
In principle, higher curvature terms are also allowed.

Integrating out the auxiliary field $e_{\m\n}$, we have a Lagrangian of higher curvature gravity, which schematically reads 
\be
\mathcal L=\sqrt{-g}\left(M_p^2\,\L+M_p^2\,R+\frac {M_p^2} {m^2}R\, R+\dots\right).
\ee
We can see a small mass in the bi-gravity formulation 
translates into large higher curvature terms. 

To be more precise, there are two kinds of Ostrogradsky's ghosts in 
a model of higher curvature gravity, which can be rephased more transparently in the bi-gravity formulation. 
In the bi-gravity representation, the massive modes contain two kinds of ghost-like degrees of freedom. 
The first one is the ghost-like spin-2 mode due to a wrong sign kinetic term
\footnote{The kinetic term of the helicity-1 mode is from the mass terms, so they are ghost when the spin-2 ghost is also a tachyon. 
Let us assume the mass squared is positive. }. 
The second one is Ostrogradsky's scalar ghost or the Boulware-Deser ghost in a generic theory of 
massive spin-2 field. 

For simplicity, let us assume Minkowski spacetime is the background solution. 
We also assume, after linearization and diagonalization, 
the kinetic terms of \eqref{bigra-gen} are given by 
the linearized Einstein-Hilbert terms to avoid more ghosts. 
These assumptions already constrain the possible terms in \eqref{bigra-gen}. 
The scale of the quadratic potential terms is set by the mass squared $m^2$. 
Then both the spin-2 ghost and the scalar ghost are excited 
and interact with the healthy modes 
at a low scale 
\be
\L=m.
\ee  

By requiring that the quadratic term of the linearized Lagrangian 
takes the form of Fierz-Pauli mass term, 
the scalar ghost is absent in the quadratic action. 
But it can still appear in the interaction terms, which is known as the BD ghost. 
If we assume the effective Lagrangian is given by 
the bi-gravity models in the second class \eqref{2cl-2}, 
this scalar ghost can be eliminated completely. 
Then we can focus on the problem of spin-2 ghosts. 

To increase the cut-off scale, let us first examine 
the quadratic action in detail. 
The massive spin-2 field is denoted by $H_{\m\n}$. 
According to section \ref{sec-lin}, 
its linearized action is given by Fierz-Pauli theory  
\ba
\bar{\mathcal L}&=&
M_p^2\big(H_\m{}^{[\m}\pa_\n\pa^\n H_\r{}^{\r]}
+ \a\,m^2 H_\m{}^{[\m} H_\n{}^{\n]}\big), 
\ea
where the kinetic term has a wrong sign and $\a$ is 
a model-dependent numerical factor.

Let us decompose $H_{\m\n}$ $\grave{\text{a}}$ la Helmholtz
\be
H_{\m\n}=\frac 1 {M_p}H^T_{\m\n}
+\frac 1 {M_p\, m}(\pa_\m A_\n+\pa_\n A_\m)
+\frac 1 {M_p\, m^2}\pa_\m \pa_\n \phi,
\ee
\be
\pa^\m H^T_{\m\n}=0,\quad\pa^\m A_\m=0.
\ee
where the dimensions of $H^T_{\m\n},\,A_\m,\,\phi$ are 1 and 
the use of $M_p,\, m$ is to canonically normalize the kinetic terms 
of the decomposed fields
\footnote{Another natural decomposition is with respect to the covariant derivative
of the massless spin-2 field $h_{\m\n}$
\be
H_{\m\n}=\frac 1 {M_p}H^T_{\m\n}
+\frac 1 {M_p\, m}(\nabla^{(h)}_\m A_\n+\nabla^{(h)}_\n A_\m)
+\frac 1 {M_p\, m^2}\nabla^{(h)}_\m \pa_\n \phi
\ee
\be
{\nabla^{(h)}}^\m H^T_{\m\n}=0,\quad 
{\nabla^{(h)}}^\m A_\m=0.
\ee
}.

In terms of the decomposed modes, 
the Fierz-Pauli Lagrangian becomes
\ba
\mathcal L=&&\,
\pa_\m {(H^T)_{\n\r}} \pa^\m {(H^T)^{\n\r}}
-\,\pa_\m {(H^T)_\n}^\n \pa^\m {(H^T)_\r}^\r
\nonumber\\&&\,
+\a\,m^2\,\left[{(H^T)_\m}^\m{(H^T)_\n}^\n
-\,(H^T)_{\m\n} (H^T)^{\m\n}\right]
\nonumber\\&&\,
-2\,\a\,\pa_\m A_\n\, \pa^\m A^\n
+2\,\a\, {(H^T)_\m}^\m(\Box\Phi),\qquad
\ea
where total derivative terms are neglected. 
The last term is a cross term, 
so we introduce
\be
\bar H_{\m\n}=H^T_{\m\n}+\frac {\a}3 \phi\, \eta_{\m\n}
\ee
to diagonalize the kinetic terms. 
After diagonalization, 
the Lagrangian becomes
\ba
\mathcal L=&&\,
\pa_\m {{\bar H}_{\n\r}} \pa^\m {{\bar H}^{\n\r}}
-\,\pa_\m {{\bar H}_\r}{}^\r \pa^\m {{\bar H}_\n}{}^\n
\nonumber\\&&\,
+\a\,m^2\,\big({{\bar H}_\m}{}^\m{{\bar H}_\n}{}^\n
-\,{\bar H}_{\m\n} {\bar H}^{\m\n}\big)
\nn&&
+\frac 4 3 \a^2 \pa_\m\phi\,\pa^\m\phi
+2\a^2\,m^2{\bar H}_\m{}^\m\phi
+\frac 4 3 \a^3 m^2\phi^2
\nonumber\\&&\,
-2\,\a\,\pa_\m A_\n\, \pa^\m A^\n.
\ea
The kinetic terms of $\bar H_{\m\n}$ and $\phi$ have wrong signs.  
If $\a>0$, then the helicity-1 modes $A_\m$ are healthy modes. 
Although the ghost-like modes can be excited at scale $\L=m$, 
they are harmless 
before healthy degrees of freedom are coupled to them. 
The cut-off scales are then determined by 
the lowest scale of the interaction terms 
that involve both the ghosts and the healthy modes. 

By considering nonlinear redefinitions, 
we can always make the massless spin-2 field transverse
\be
h_{\m\n}=\eta_{\m\n}+\frac 1 {M_p} h^T_{\m\n}.
\ee

Then we perform a general power counting of 
the possible perturbative terms 
without using the specific structures of the nonlinear terms. 
From the two derivative terms, we have
\ba
M_p^{2-i-2j-k-l} m^{-2j-2k} \pa^2 (H^T)^{i} (\pa A)^{2j} (\pa\pa\phi)^k (h^T)^l.
\nn
\ea

Assuming the coefficients of the potential terms in \eqref{bigra-gen} are of the same order,
we have
\ba
M_p^{2-i-2j-k-l} m^{2-2j-2k} (H^T)^i (\pa A)^{2j} (\pa\pa\phi)^k (h^T)^l.
\nn
\ea

The interaction terms 
start from the cubic order
\be
i+2j+k+l=3,\,4,\,5,\dots, 
\ee
and a perturbative term 
\be
M_p^{-a}\, m^{-b}  \pa^m (H^T)^{i} (A)^{2j} (\phi)^k (h^T)^l.
\ee
becomes important at the energy scale
\be
\L=(M_p^a m^b)^{1/(a+b)}.
\ee

The lowest scales of the interaction terms involving both the ghost-like modes 
$(H^T,\,\phi)$ and the healthy modes $(h^T,\,A)$
\footnote{If $\a<0$, then the helicity-1 mode $A_\m$ becomes 
a ghost due to a wrong sign kinetic term. 
If $a=0$, the helicity-1 mode is strongly coupled. 
In addition, $A_\m$ could be Ostrogradsky's vector ghost 
if the equations of motion involve higher order time derivative terms of $A_\m$. 
But from the constraint analysis \cite{Li:2015iwc}, 
we can count the numbers of dynamical degrees of freedom, 
and we know the novel kinetic terms do not contain Ostrogradsky's vector ghost. 
} 
can be found:
\begin{itemize}
\item
The lowest scale of the two-derivative terms is
\be
\L_5=(M_p\, m^4)^{1/5},
\ee
where the cubic terms
\be
j=k=1,\quad i=l=0,
\ee
and
\be
k=2,\quad l=1,\quad i=j=0
\ee
become important. 

\item
The lowest scale of the potential terms is
\be
\L_3=(M_p\, m^2)^{1/3},
\ee
where infinitely many terms
\be
j=1,\quad k=1+n,\quad i=l=0
\ee
and
\be
k=2+n,\quad l=1,\quad i=j=0
\ee
\be
n=0,\,1,\,2,\,\dots
\ee
become important. 

\end{itemize}

The cut-off scale of the two-derivative terms is lower than 
that of the potential terms, 
so a generic bi-gravity model in the second class \eqref{2cl-2} 
is a consistent effective field theory at least up to $\L_5$. 
\footnote{If the potential terms are modified, then
Ostrogradsky's scalar ghost is eliminated only at the linear level. 
If the massive spin-2 field has a correct sign kinetic term, 
then the cut-off scale set by the Boulware-Deser ghost is $\L_5$ as well. 
} 

We can further improve this by turning off $\mathcal L_3^{\text{kin}}$, 
then the kinetic terms have more gauge symmetries.  
The kinetic terms contain only helicity-2 modes  
whose the interaction scale is set by the Planck mass. 
In this way, we are able to increase the cut-off scale to at least $\L_3$
\footnote{In \cite{Hinterbichler:2015soa}, it was shown that 
the $\L_3$ interaction terms vanish in some cases 
which are equivalent to quadratic curvature gravity. 
This indicates the cut-off scale may be higher than $\L_3$. }
\be
\L\rightarrow \L_3\equiv (M_p m^2)^{1/3}.
\ee 

It should be noted that 
we need to make sure the spin-2 ghost does not couple to the matter 
below the cut-off scale. 
This indicates we should consider an effective metric \cite{eff-metric}
\footnote{But the BD ghost will be revived  at some scale above $\L_3$ due to the use of an effective metric. }. 
\\

In a different region of parameter space, 
it is possible that the massless spin-2 field has a wrong sign, 
while the massive one is healthy. 
By eliminating the BD-ghost and using the gauge-invariant kinetic terms, 
the cut-off scale is set by 
the interaction terms $k=2+n,\,l=1,\,i=j=0$ from the potentials, 
which is $\L_3$ again. 
When the mass squared has a correct sign, 
these effective field theory contain a ghost-free massive graviton and 
a decoupled ghost-like, massless spin-2 field 
below the cut-off scale. 
\\

Here we want to give one possible reason 
for ``naturally" large coefficients of the higher curvature terms. 
In the bi-gravity formulation, 
the gauge symmetries are enhanced when 
\be
m^2=0,\quad a_3=0, 
\ee
so small values of $m^2$ and $a_3$ are technically natural, 
which is analogous to the mass of electron. 
Quantum corrections to these parameters 
should be multiplicative, rather than additive.  

\section{Symmetric condition}\label{sec-sym}
In the above sections, 
we impose the symmetric condition or 
the Deser-van Nieuwenhuizen condition \cite{Deser:1974cy} 
to minimalize the numbers of dynamical degrees of freedom. 
In this section, we present a general way to derive 
the symmetric condition from the equations of motion. 

Let us decompose the rank-2 tensor $e_{\m\n}$ into two parts
\be
e_{\m\n}=F_\m{}^A E_\n{}^B \eta_{AB}
=e^{\text{s}}_{\m\n}+e^{\text{as}}_{\m\n}.
\ee
where $e^{\text{s}}_{\m\n}$ is the symmetric part 
and $e^{\text{as}}_{\m\n}$ is the antisymmetric part 
\be
e^{\text{s}}_{\m\n}=e^{\text{s}}_{\n\m},
\quad
e^{\text{as}}_{\m\n}=-e^{\text{as}}_{\n\m}. 
\ee

In the Lagrangians, an antisymmetric product vanishes 
if it contains an odd number of $e_{\m\n}^{\text{as}}$. 
In 4d bi-gravity models, the antisymmetric part of $e_{\m\n}$ only appears in the terms below
\be
R_{\m\n}{}^{[\m\n}(e^{\text{as}})_\r{}^{\r}(e^{\text{as}})_\s{}^{\s]},
\ee
\be
(e^{\text{as}})_\m{}^{[\m}(e^{\text{as}})_\n{}^{\n]},
\quad
(e^{\text{as}})_\m{}^{[\m}(e^{\text{as}})_\n{}^{\n}(e^{\text{s}})_\r{}^{\r]},
\ee
\be
(e^{\text{as}})_\m{}^{[\m}(e^{\text{as}})_\n{}^{\n}(e^{\text{s}})_\r{}^{\r}(e^{\text{s}})_\s{}^{\s]},
\ee
\be
(e^{\text{as}})_\m{}^{[\m}(e^{\text{as}})_\n{}^{\n}(e^{\text{as}})_\r{}^{\r}(e^{\text{as}})_\s{}^{\s]}. 
\label{4-anti-e}
\ee

We argue that the equations of motion for $e_{\m\n}^{\text{as}}$ generally lead to the symmetric condition
\be
\frac {\d}{\d e_{\m\n}^{\text{as}}}\int\mathcal L=0
\quad\Rightarrow 
\quad
e_{\m\n}^{\text{as}}=0,
\ee
because the equations for $e_{\m\n}^{\text{as}}$ can be written in a matrix form
\be
\mathcal A_{\m\n,\r\s} (e^{\text{as}})^{\r\s}=0, 
\ee
which gives the symmetric condition if $\mathcal A$ is invertible.  
An important point is that the Lagrangians do not contain linear terms, 
so the equations of motion for $e^\text{as}$ are homogeneous
\footnote{We assume the matter does not couple to the antisymmetric part of $e_{\m\n}$ linearly. 
For example, 
if the physical vielbein is a linear combination of $E_\m{}^A$ and $F_\m{}^B$, 
the corresponding physical metric will contain a quadratic term 
$e^{\text{as}}_{\m\r}e^{\text{as}}_\n{}^\r$, 
so the equations of motion for $e^{\text{as}}_{\m\n}$ are homogeneous. }. 

The argument is clear if \eqref{4-anti-e} is not considered. 
When the Lagrangian contains \eqref{4-anti-e}, 
the equations of motion contain cubic terms of $e_{\m\n}^{\text{as}}$.  
Then we can write the cubic terms as products of quadratic terms and linear terms, 
and think of the quadratic terms as part of the matrix $\mathcal A$. 
\footnote{Note that $\mathcal A$ can be degenerate for special values of $e_{\m\n}^{\text{as}}$ 
if they are not the solutions of the equations of motion at the same time. }

We do not have a proof that the above argument works in general, 
but we check several examples and always find that 
\be
\det \mathcal A\neq 0. 
\ee 
The spirit is close to \cite{Hinterbichler:2012cn} , 
where the symmetric condition is derived from a local Lorentz transformation. 
In addition, we do not rule out the possibility that 
$\mathcal A$ could be degenerate at some singular points of the phase space. 

\section{Higher derivative generalizations}\label{sec-highder}
Along the line of Lovelock terms, 
the novel kinetic terms can be generalized to novel higher-derivative terms 
\cite{gh-fr}
\footnote{If we impose the symmetric condition 
and fix the second metric to be Minkowski, 
they are equivalent to 
the higher derivative interactions 
proposed in \cite {Kimura:2013ika}.}
\be
R(E)\wedge\dots \wedge R(E)\wedge E\wedge \dots \wedge E \wedge F\wedge \dots \wedge F, 
\label{higher-derivative-bigra}
\ee
which might be inconsistent with terms involving $R(F)$. 

In section \ref{sec-L2}, we discuss the special cases with only one $F$ vielbein
\be
R(E)\wedge\dots \wedge R(E)\wedge E\wedge \dots \wedge E \wedge F,
\label{1-F}
\ee 
which have the same gauge symmetries as $R(E)\wedge E \wedge F$. 
They describe the derivative interactions between two massless, gauge invariant spin-2 fields, 
where the Boulware-Deser ghost is absent. 
We expect other higher derivative terms in \eqref{higher-derivative-bigra} do not contain the BD ghost as well.

For the extension to multi-gravity, we have 
\be
R(E^{(1)})\wedge\dots \wedge R(E^{(1)})\wedge E^{(2)}\wedge \dots \wedge E^{(d-n)}, 
\label{higher-der}
\ee
where $n$ is the number of curvature 2-forms and 
$E^{(k)}$ vielbeins can be the same or different. 
Lovelock terms and dRGT terms are unified in \eqref{higher-der}. 
\\

In the end, we would like to connect with some results in the literature.  
The bi-gravity models with \eqref{1-F} in the metric formulation 
were already proposed in \cite{Paulos:2012xe}
as generalizations of New Massive Gravity. 
The BD ghost was argued to be absent 
by counting the degrees of freedom using symmetries \cite{deRham:2011ca, Paulos:2012xe}. 
The antisymmetric structure guarantees that the equations of motion for the decomposed fields 
\be
e_{\m\n}=\tilde e_{\m\n}+\nabla_\m A_\n+\nabla_\n A_\m+\nabla_\m\pa_\n \phi
\label{decomposed}
\ee
are of second order, 
so additional degrees of freedom are avoided
\footnote{This argument is dangerous. 
The equation of motion for the decomposed field $\phi$ 
contains third order derivative terms of the metric, 
in the form of covariant derivatives of curvature tensors. 
In addition, if one varies the action with respect to the metric after the substitution, 
the equations of motion will contain third order derivative terms of $\phi$ 
due to the variations of covariant derivatives. 
But it is possible that the third order time derivative terms can be removed by 
the time derivatives of some second order equations \cite{Deffayet:2015qwa}, 
then the counting of the degrees of freedom is correct.
}. 
Then one can count the dynamical degrees of freedom in the bi-gravity models 
and show that the total number is at most $(d^2-2d-1)$, 
which is that of a massless and a massive spin-2 fields. 
Therefore, the Boulware-Deser ghost is absent
\footnote{When there are additional gauge symmetries, 
the number of dynamical degrees of freedom is reduced.}. 

From this argument, 
we can see why the curvature tensors should be associated with the same spin-2 field $g_{\m\n}$. 
If a curvature tensor contains the second spin-2 field $e_{\m\n}$, 
then the equations of motion for the decomposed modes will usually be of higher order, 
because they are not gauge modes in a curvature tensor 
and no apparent antisymmetric structure is protecting them
\footnote{In \cite{Akhavan:2016hju}, an Einstein-Hilbert term for $e_{\m\n}$ was introduced 
to obtain unitary models. However, we suspect the absence of ghost-like degrees of freedom is an artifact of linearization. For example, in the minisuperspace approximation, the Hamiltonians are not linear in the lapse functions. }.
\\

The decomposed field argument concerning \eqref{1-F} is based on the fact that 
Lovelock tensors are divergence-free. 
For the other novel derivative terms in \eqref{higher-der}, 
we can generalize this argument by using the second Bianchi identity
\be
\nabla_{[\m} R_{\n\r]}{}^{\a\b}=0.
\ee
The covariant derivatives in front of the decomposed fields will not act on the Riemann tensor after integrating by parts, 
so the equations of motion for the decomposed fields will not contain 4th order derivative terms of the metric. 
The variation of a Riemann tensor $R_{\m\n}{}^{\r\s}$ contains 
some second covariant derivative  terms of $\d g_\m{}^\n$ which are antisymmetrized , 
so the equations of motion for the metric will not contain 4th order derivative terms of $\phi$. 

In the vielbein formulation, 
the second Bianchi identity stems from a basic identity of exterior derivative
\be
d^2=0,
\ee
which is the key element of the unifying framework \cite{gh-fr, Li:2015fxa}. 

\section{Conclusion}\label{sec-sum}
In summary, we present evidence that
\be
R(E^{(1)})\wedge\dots\wedge
R(E^{(1)})\wedge
E^{(2)}\wedge\dots \wedge E^{(n)}
\label{sum-novel-der}
\ee
are basic building blocks for the actions of interacting spin-2 fields 
that are free of the Boulware-Deser ghost
\footnote{Parity is assumed to be preserved, 
otherwise there are more possible terms.  
For example, in 3d, one could introduce 
a gravitational Chern-Simons term 
which violates parity \cite{Deser:1981wh}. 
The critical points of higher derivative gravity theories were first investigated in this context \cite{Li:2008dq}.}. 
Models that can be constructed from these building blocks 
include Einstein gravity, Weyl gravity, Lovelock gravity, 
New Massive Gravity, dRGT massive gravity and 
some of their generalizations. 
The parameter space is further extended by novel derivative terms. 

Curiously, the building blocks \eqref{sum-novel-der} 
can be obtained from Lovelock terms by replacing  
some of the vielbeins in the wedge products with other vielbeins. 
\\

The novel two-derivative terms in 4d are studied in detail:
\begin{itemize}
\item
Based on a minisuperspace analysis, 
a large class of bi-gravity models \eqref{2cl-2} are identified, 
which are potentially free of the Boulware-Deser ghost. 

\item
The bi-gravity models in this class \eqref{2cl-2} do not have 
the usual single dynamical metric limit with a fixed metric, 
which is in accordance with 
the no-go theorem for new kinetic interaction for 
single dynamical metric in \cite{deRham:2013tfa}. 

\item 
We reformulate some well-understood models of higher curvature gravity as bi-gravity models in this class \eqref{2cl-2}, 
Their spectra are known to contain 
1 massless and 1 massive spin-2 fields 
without the BD ghost. 

\item
The argument that New Massive Gravity is free of the BD ghost 
is extended to other bi-gravity models in this class \eqref{2cl-2}, 
which applies to novel higher-derivative terms as well 
\footnote{The absence of the BD ghost 
in 4d novel kinetic terms is proved by the constraint analyses 
in \cite{Li:2015iwc}.}.
\end{itemize}
 
This class of bi-gravity models are interesting 
despite the issue of spin-2 ghosts. 
Firstly, as toy models of quantum gravity, 
they have better chance to be perturbative renormalizable 
and there are less negative norm states 
because the BD-ghost is absent. 
Secondly, as effective field theories of gravity, 
they can increase the cut-off scale 
set by higher derivative terms with large coefficients
\footnote{An important question is 
whether the special structure of BD-ghost-free building blocks 
are detuned by quantum corrections. 
}. 

In general, we can avoid the ghost modes 
by reducing the number of dynamical degrees of freedom. 
A useful strategy of eliminating ghost modes 
is to impose specific boundary conditions. 
Another method to remove the spin-2 ghosts is simply setting the decomposed helicity-2 modes $\tilde e_{\m\n}$ in \eqref{decomposed} to zero. 
They may give rise to healthy vector-tensor theories
\footnote{We refer to \cite{vec-ten} for recent developments on vector-tensor models,}.
\\

In the light of the AdS/CFT correspondence \cite{holo}, 
the large class of non-unitary bi-gravity models 
may be dual to non-unitary conformal field theories. 
It is interesting to explore non-unitary holography 
in the extended parameter space. 

\begin{acknowledgments}
Acknowledgment: 
I would like to give special thanks to Xian Gao, Elias Kiritsis and Ryo Saito for inspiring discussions. 
I am grateful to E. Babichev, C. Charmousis, E. Joung, J. Mourad, V. Niarchos, F. Nitti, K. Noui and D. Steer for useful comments or/and discussions. 
I want to thank C. de Rham, K. Hinterbichler, A. Matas, A. Solomon and A. Tolley for correspondence.
I also thank A. Tseytlin for pointing out a missed reference. 

This work was supported in part by European Union's Seventh Framework Programme
under grant agreements (FP7-REGPOT-2012-2013-1) no 316165, the EU program ``Thales" MIS 375734
 and was also cofinanced by the European Union (European Social Fund, ESF) and Greek national funds through
the Operational Program ``Education and Lifelong Learning" of the National Strategic
Reference Framework (NSRF) under ``Funding of proposals that have received
a positive evaluation in the 3rd and 4th Call of ERC Grant Schemes".
\end{acknowledgments}

\end{document}